\documentclass{article}

\usepackage{geometry}
\usepackage{calc}
\usepackage{amsmath}
\usepackage{amssymb}
\usepackage{stmaryrd}
\usepackage{graphicx}
\usepackage{subfigure}

\usepackage{particles}
\usepackage{numinsec}
\usepackage{pptitle}

\title{A Note on One Loop Electroweak Contributions to $g-2$: a
  Companion to \textsc{BUHEP-01-16}}
\author{Kevin R. Lynch\thanks{krlynch@bu.edu}\\ \\
Department of Physics\\ Boston University\\ 590 Commonwealth Avenue\\
Boston, MA 02215}
\date{August 7, 2001}
\preprintwidth{hep-ph/0108081}
\preprints{BUHEP-01-17\\hep-ph/0108081}

\newcommand{\amuon}{\ensuremath{{a_\muon}}}
\newcommand{\afermion}{\ensuremath{{a_\particle{f}}}}
\newcommand{\aext}{\ensuremath{{a_\external}}}

\newcommand{\Psif}{{\ensuremath{{\particle{\Psi}_\particle{f}}}}}
\newcommand{\Psifbar}{{\ensuremath{{\DB{\particle{\Psi}}_\particle{f}}}}}

\newcommand{\dd}[2][]{\ensuremath{{\mathrm{d}^{#1}{#2}}}}

\newcommand{\Spinorubar}{\ensuremath{{\overline{\mathrm{u}}(p')}}}
\newcommand{\Spinoru}{\ensuremath{{\mathrm{u}(p)}}}
\newcommand{\Spinorubark}{\ensuremath{{\overline{\mathrm{u}}(k')}}}
\newcommand{\Spinoruk}{\ensuremath{{\mathrm{u}(k)}}}

\newcommand{\amplitude}{\ensuremath{{\mathcal{M}}}}
\newcommand{\Acl}{\ensuremath{{\tilde{A}^{\text{cl}}_\mu(q)}}}

\newcommand{\cphi}{\ensuremath{{c_\phi}}}
\newcommand{\sphi}{\ensuremath{{s_\phi}}}

\newcommand{\internal}{{\text{int}}}
\newcommand{\external}{{\text{ext}}}

\makeatletter
\newsavebox{\@pslash}
\savebox{\@pslash}{\ensuremath{{\DiracSlash{p}}}}
\newcommand{\pslash}{\usebox{\@pslash}}

\newsavebox{\@ppslash}
\savebox{\@ppslash}{\ensuremath{{\smash[t]{\DiracSlash{p}}'}}}
\newcommand{\ppslash}{\usebox{\@ppslash}}

\newsavebox{\@kslash}
\savebox{\@kslash}{\ensuremath{{\DiracSlash{k}}}}
\newcommand{\kslash}{\usebox{\@kslash}}

\newsavebox{\@kpslash}
\savebox{\@kpslash}{\ensuremath{{\smash[t]{\DiracSlash{k}}'}}}
\newcommand{\kpslash}{\usebox{\@kpslash}}

\newsavebox{\@qslash}
\savebox{\@qslash}{\ensuremath{{\DiracSlash{q}}}}
\newcommand{\qslash}{\usebox{\@qslash}}

\newsavebox{\@lslash}
\savebox{\@lslash}{\ensuremath{{\DiracSlash{\ell}}}}
\newcommand{\lslash}{\usebox{\@lslash}}
\makeatother

\newcommand{\phivev}{\ensuremath{{\left<\phi\right>}}}
\newcommand{\phisvev}{\ensuremath{{\left<\phi^*\right>}}}

\begin{document}

\begin{titlepage}
\maketitle
\begin{abstract}
  In this note we present general expressions at one loop order that
  can be used to calculate the contributions to the anomalous magnetic
  moment of fundamental, charged Dirac fermions.  In particular, we
  provide the expressions for charged and neutral scalar and charged
  and neutral gauge boson contributions with general vector and axial
  couplings to the fermion of interest.  The calculations presented in
  this note were originally derived for use in the author's letter,
  \cite{gauge}.  We have chosen to document and make available the
  results and derivations in the hope that they will also be useful to
  others.  Our expressions reproduce the Standard Model electroweak
  contributions to \amuon\ in the appropriate mass limits, and are
  flexible enough to allow us to handle many scenarios of new physics
  beyond the Standard Model.
\end{abstract}
\end{titlepage}

\section{Introduction}
\label{sec:intro}

The recent measurement of the anomalous magnetic moment of the muon,
\amuon, by the Brookhaven E821 Collaboration \cite{Brown:2001mg},
showing a possibly significant discrepancy between experimental
measurement and the Standard Model theoretical expectation, has
generated substantial new interest in the field.  In this note we
present general expressions at one loop order that can be used to
calculate the contributions to the anomalous magnetic moment of
fundamental, charged Dirac fermions.  In particular, we provide the
expressions for charged and neutral scalar and charged and neutral
gauge boson contributions with general vector and axial couplings to
the fermion of interest.  By ``general'', we mean that we have
explicitly allowed permitted, for instance, the possibility of
tree-level flavor changing and $CP$ violating couplings.

The calculations presented in this note were originally derived for
use in the author's letter, \cite{gauge}.  Other authors have
presented results with a similar physics motivation; in particular we
cite the first calculation of the weak contributions
\cite{Jackiw:1972jz}, the first calculation in general renormalizable
gauges \cite{fls} and the first calculation for general gauge models
\cite{Leveille:1978rc}.  The main difference between earlier works and
this work is that we take a pedagogical approach whenever possible, in
the hope that they will be helpful to those learning the techniques
necessary for these types of calculations.

In the next section, we construct the building blocks, or tools, for
this calculation.  In particular, we provide a careful definition of
our conventions, derivations of the Feynman rules needed in the
calculations to come, derivations of kinematic factors, and discussion
of the numerous Dirac algebra relations that arise in the numerators
of the amplitudes.  In Section~\ref{sec:building} we use the tools we
have built to derive some very general expressions for the
contributions of neutral and charged scalars and vectors to the
anomalous magnetic moment of the charged fermions.  We perform all
gauge calculations in Feynman Gauge in the narrow width approximation;
removing these restrictions, in particular the narrow width
assumption, would be the most fruitful addition to the results of this
note. In Section~\ref{sec:SM}, we apply those very results to obtain
expressions in the context of the Standard Model for charged leptons,
and in particular the muon.  In Section~\ref{sec:nonSM}, we look
further at some simple applications to potential physics contributing
to \amuon\ arising in extensions to the Standard Model.  Finally,
Section~\ref{sec:conclusion} provides some conclusions and directions
for extending this work.

\section{A Systematic Approach: Constructing the Tools}
\label{sec:tools}

In this section, we first review our conventions, and then turn to a
number of subcalculations that are necessary for the calculation of
the anomalous magnetic moment contributions.  For a general overview,
of course, one should first peruse a standard Quantum Field Theory
textbook, such as Peskin and Schroeder \cite{peskin}.

\subsection{Definition of the Form Factors}

\begin{figure}
\begin{center}
\includegraphics[width=(\textwidth/2-1in)]{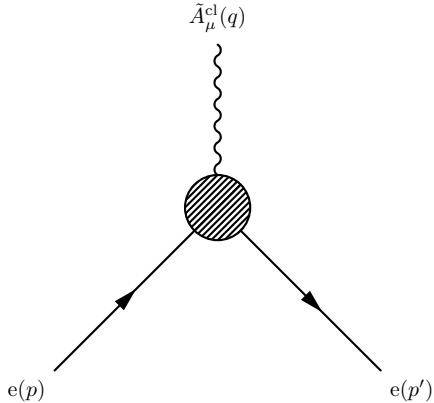}
\caption{A ``Feynman Diagram'' cartoon for lepton scattering
  (specifically, electron) from a static, classical background of
  photons, \Acl.  The shaded circle is a stand-in for the full vertex
  function.}
\label{fig:full-lepton-vertex}
\end{center}
\end{figure}

The ``Feynman diagram'' for lepton-photon scattering is shown in
Figure~\ref{fig:full-lepton-vertex}.  The amplitude for leptons
scattering off a static (classical) background field is given by 
\begin{equation}
i \amplitude = -ieQ_\lepton \Acl \Spinorubar
\Gamma^\mu \Spinoru\ ,
\label{eqn:amplitude}
\end{equation}
where the vertex operator $\Gamma^\mu$ can only be a function of
$p^\mu$, ${p'}^\mu$, $q^\mu$, \DiracGamma{\mu}, and \DiracGamma{5}.
Not all combinations are independent (the Gordon identity, for
example, links them).  The conventional combination is written
\begin{equation}
\Gamma^\mu = F_1(q^2) \DiracGamma{\mu} + F_2(q^2) \frac{i
  \sigma^{\mu\nu} q_\nu}{2m_\lepton} + F_3(q^2) \frac{i
  \sigma^{\mu\nu} q_\nu \DiracGamma{5}}{2m}\ .
\label{eqn:definition}
\end{equation}
The terms are chosen in this way because, in the limit $q^2 \to 0$
(\textit{i.e.} when all external particles are put on shell), the
functions $F_i$ correspond to classical definitions of the electric
charge ($F_1(0)$), anomalous magnetic moment ($F_2(0)$), and electric
dipole moment ($F_3(0)$).  Although we will not need to consider these
issues, note that $F_1(0)$ is constrained to be 1, and thus enters as
a renormalization condition.  Further, $F_3(0)$ is strictly zero in
QED, and we will not consider it further at this time.  Finally, we
will find that all of the integrals in the loop diagrams that give us
$F_2(q^2)$ are convergent, and so we will not need to involve
ourselves at all in the renormalization program, although at higher
loop order, this is no longer the case.  To get the notation straight,
let us ``calculate'' the $F_i$ at tree level in QED.  The amplitude
for an electron scattering from a static photon background is given by
\begin{equation}
i \amplitude = \Spinorubar \left( -ieQ_\lepton \DiracGamma{\mu}
\right) \Spinoru \Acl\ .
\end{equation}
Clearly, when we rearrange this as in Equation~\ref{eqn:amplitude}, we 
find $F_1(q^2) = 1$ and $F_2(q^2) = F_3(q^2) = 0$.  

Note that in this work the anomalous magnetic moment of the fermion
\particle{f}\ corresponds to
\begin{equation}
\afermion = \frac{g_\particle{f}-2}{2} = F_2(0)\ ,
\end{equation}
which can be compared directly to the results of the E821
collaboration.

\subsection{Lagrangians and Feynman Rules}

We take the following as our fermionic Lagrangian
\begin{equation}
\mathcal{L} = \Psifbar \left(i \DS{D} - m_\Psif \right)\Psif = \Psifbar
\left( i \DG{\mu} \left(\partial_\mu - i g
    \particle{A}^\mathrm{a}_\mu t^\mathrm{a} \right) -
  m_\Psif\right) \Psif\ ,
\end{equation}
which gives us the following Feynman rule for gauge-fermion coupling
\begin{equation}
i g \DG{\mu} t^\mathrm{a}\ .
\end{equation}

The Feynman rules for scalar fields and gauge fields are more
complicated to derive, and in fact, many texts with seemingly similar
conventions disagree on the derived Feynman Rules; additionally, at
least one text (Cheng and Li \cite{cheng}) contains internally
inconsistent results for some amplitudes.  Although this disagreement
is limited to factors of $i$ and $-1$, which does not matter when
calculating most observables as they arise from squared amplitudes,
these factors become very important when the observable being
calculated is contained in directly in the amplitude, as is the
case here.

\subsubsection{Unbroken Gauge Theory}

Let us consider first the Feynman rules for an unbroken gauge theory.
The Lagrangian for an unbroken, non-abelian gauge theory
\begin{equation}
\mathcal{L} = -\frac{1}{4} {F_{\mu\nu}^\mathrm{a}}^2 - \frac{1}{2\xi}
\left(\partial_\mu A^{\mu\mathrm{a}}\right)^2\ ,
\end{equation}
where the first term is
\begin{equation*}
F_{\mu\nu}^\mathrm{a} = \partial_\mu A_\nu^\mathrm{a} - \partial_\nu
A_\mu^\mathrm{a} + g f^{\mathrm{abc}} A_\mu^\mathrm{b}
A_\nu^\mathrm{c}\ ,
\end{equation*}
and the second term is the Fadeev-Popov gauge fixing term, and (in
this case) affects only the propagator.  If we expand this term to
find the trilinear and quadrilinear terms, we find (after defining
$f_{\mu\nu}^\mathrm{a} = \partial_\mu A_\nu^\mathrm{a} - \partial_\nu
A_\mu^\mathrm{a}$),
\begin{multline*}
F_{\mu\nu}^\mathrm{a} F^{\mu\nu\mathrm{a}} = f_{\mu\nu}^\mathrm{a}
f^{\mu\nu\mathrm{a}}\\
+\left(\partial_\mu A_\nu^\mathrm{a} - \partial_\nu
A_\mu^\mathrm{a}\right) g f^{\mathrm{abc}} A^{\mu\mathrm{b}}
A^{\nu\mathrm{c}} + g f^{\mathrm{ade}} A_\mu^\mathrm{d}
A_\nu^\mathrm{e} \left(\partial^\mu A^{\nu\mathrm{a}} - \partial^\nu
A^{\mu\mathrm{a}}\right)\\
+ g^2 f^{\mathrm{abc}} f^{\mathrm{ade}} A_\mu^\mathrm{b}
A_\nu^\mathrm{c} A^{\mu\mathrm{d}} A^{\nu\mathrm{e}}\ .
\end{multline*}
The first line contributes to the propagator, the second line
contributes to the trilinear gauge coupling, and the third line
contributes to the quadrilinear couplings.  

Let us start with the trilinear term, and derive it in detail, since
various texts with supposedly identical conventions obtain results
differing by factors of $i$.  We need first some conventions that all
the texts agree on.  We label external initial state particles with
incoming momenta, and external final state particles with outgoing
momenta. For an external initial state gauge boson, we assign a factor 
\begin{equation*}
A_\mu(p) = \varepsilon_\mu(p) e^{-ipx} \to \partial_\mu A_\nu = -i
p_\mu A_\nu\ ,
\end{equation*}
and for an external final state gauge boson, we assign a factor
\begin{equation*}
A_\nu(p') = \varepsilon^*_\nu(p') e^{ip'x} \to \partial_\mu A_\nu =
ip'_\mu A_\nu\ .
\end{equation*}
Then, the trilinear term in $\mathcal{L}$, with all particles
considered ``initial state'' (i.e. with incoming momenta), is given by
\begin{gather*}
\mathcal{L} \supset -\frac{1}{2} f^{\mathrm{abc}} \left(\partial_\mu
  A_\nu^\mathrm{a}\right) A^{\mu\mathrm{b}} A^{\nu\mathrm{c}}
+\frac{1}{2} f^{\mathrm{abc}} \left(\partial_\nu
  A_\mu^\mathrm{a}\right) A^{\mu\mathrm{b}} A^{\nu\mathrm{c}}\ . \\
\intertext{In the second term, we swap the dummy variables $\mu \leftrightarrow
  \nu$ and $b \leftrightarrow c$}
\mathcal{L} \supset -\frac{1}{2} f^{\mathrm{abc}} \left(\partial_\mu
  A_\nu^\mathrm{a}\right) A^{\mu\mathrm{b}} A^{\nu\mathrm{c}}
+\frac{1}{2} f^{\mathrm{acb}} \left(\partial_\mu
  A_\nu^\mathrm{a}\right) A^{\nu\mathrm{c}} A^{\mu\mathrm{b}}\ . \\
\intertext{Now, since $f^{\mathrm{abc}}$ is completely antisymmetric,
  we make the substitution $f^{\mathrm{acb}} \to -f^{\mathrm{abc}}$,
  and obtain}
 - gf^{\mathrm{abc}} \left(\partial_\mu A_\nu^\mathrm{a}\right)
A^{\mu\mathrm{b}} A^{\nu\mathrm{c}} \\
= ig f^{\mathrm{abc}} k_\mu g^{\alpha\mu} g^{\beta\nu}
A_\nu^\mathrm{a} A_\alpha^\mathrm{b} A_\beta^\mathrm{c}\ .
\end{gather*}
From this, we obtain a ``preliminary'' Feynman rule by multiplying by $i$
\begin{equation*}
- g f^{\mathrm{abc}} k^\alpha g^{\beta\nu}\ .
\end{equation*}
Since there are $3! = 6$ ways to connect the $A_\mu^\mathrm{a}$ to a
trilinear vertex. At this point, to compare with the notation of
\cite{peskin} and \cite{cheng}, we relabel this to
\begin{equation*}
- g f^{\mathrm{abc}} k^\rho g^{\sigma\nu}\ .
\end{equation*}
Summing these terms using the total antisymmetry of
$f^{\mathrm{abc}}$, we obtain the Feynman rule
\begin{equation}
g f^{\mathrm{abc}} \left[ g^{\sigma\nu} (q-k)^\rho + g^{\rho\nu}
  (k-p)^\sigma g^{\rho\sigma} (p-q)^\nu\right]\ ,
\end{equation}
which agrees with the results of \cite{peskin}, but does not agree
with \cite{cheng}, which has an additional $i$.

Similar manipulations yield the quartic coupling Feynman rule, which
agrees with the results of both \cite{peskin} and \cite{cheng}
\begin{equation}
-ig^2 \left[ f^{\mathrm{abc}} f^{\mathrm{cde}} \left(g^{\mu\rho}
      g^{\nu\sigma} - g^{\mu\sigma} g^{\nu\rho} \right)
+ f^{\mathrm{ace}} f^{\mathrm{bde}} \left(g^{\mu\nu} g^{\rho\sigma} -
  g^{\mu\sigma} g^{\rho\nu} \right)
+ f^{\mathrm{ade}} f^{\mathrm{bce}} \left(g^{\mu\nu}
    g^{\rho\sigma} - g^{\mu\rho} g^{\nu\sigma} \right) \right]\ . 
\end{equation}

\subsubsection{Complex Scalar Fields}

We now approach a scalar theory coupled to an unbroken non-abelian
gauge theory, and derive the scalar-vector couplings.  We define the
covariant derivative with the same conventions as \cite{peskin}
and\cite{cheng} 
\begin{equation*}
D_\mu = \partial_\mu - ig A_\mu^\mathrm{a} t^\mathrm{a}\ .
\end{equation*}
The minus sign here results in a plus sign in the Feynman rule for
fermions, that is, $ig \DG{\mu} t^\mathrm{a}$.  The scalar Lagrangian
is
\begin{equation}
\mathcal{L} = \left(D_\mu \phi\right)^\dagger \left(D_\nu \phi\right)
g^{\mu\nu} - m_\phi \phi^\dagger \phi - \frac{\lambda}{4}
\left(\phi^\dagger \phi\right)^\dagger\ .
\end{equation}
The couplings occur in the first term on the right hand side.  Let us
expand this term and derive the Feynman rules
\begin{multline*}
\mathcal{L} \supset \left[\left(\partial_\mu -ig A_\mu^\mathrm{a}
    t^\mathrm{a}\right) \phi\right]^*_\ell \left[\left(\partial_\nu -
    ig A_\nu^\mathrm{b} t^\mathrm{b}\right) \phi\right]g^{\mu\nu} = 
\left(\partial_\mu \phi^*_\ell\right) \left(\partial_\nu \phi_m\right) 
g^{\mu\nu}\\
+ ig A_\mu^\mathrm{a} t^\mathrm{a} \phi^*_\ell \left(\partial_\nu
  \phi_m\right) g^{\mu\nu} - ig A_\nu^\mathrm{b} t^\mathrm{b}
\left(\partial_\mu \phi^*_\ell\right) \phi_m g^{\mu\nu}
+g^2 A_\mu^\mathrm{a} A_\nu^\mathrm{b} g^{\mu\nu} t^\mathrm{a}
\phi^*_\ell t^\mathrm{b} \phi_m\ ,
\end{multline*}
where the first term feeds into the propagator, the second line gives
rise to a scalar vector trilinear term, and the final line a
quadrilinear term.  First, conventions: an external scalar with
incoming momentum is a $\phi$ and $\partial_\mu \phi = -ip_\mu \phi$,
while a scalar with outgoing momentum is a $\phi^*$ and $\partial_\mu
\phi^* = ip'_\mu \phi^*$.  This gives rise to a Feynman rule for
trilinear terms of
\begin{equation}
ig \left(p'+p\right)_\mu t^\mathrm{a}_{\ell m}\ .
\end{equation}
The quadrilinear term is slightly more complicated, as there are two
ways to connect the vectors to the vertex, so we obtain a Feynman rule
\begin{equation}
ig^2 \left(t^\mathrm{a} t^\mathrm{b}\right)_{\ell m} g_{ \mu\nu} +
ig^2 \left(t^\mathrm{b} t^\mathrm{a}\right)_{\ell m} g_{ \mu\nu} =
ig^2 {\anticommutator{t^\mathrm{a}}{t^\mathrm{b}}}_{\ell m}
g_{\mu\nu}\ .
\end{equation}
In agree with \cite{zuber} and with the abelian case presented in
\cite{quigg}.  This result is not in agreement with the derivation in
\cite{cheng}, but the results of this derivation in that text are
internally inconsistent.

\subsubsection{Broken Gauge Theory}

In the more complicated case of broken gauge symmetry, we perform the
following replacement in the Lagrangian
\begin{equation*}
\phi \to \phivev + \phi\ ,
\end{equation*}
which gives rise to a ``bilinear coupling'' between $A_\mu$ and $\phi$ 
that we cancel by replacing the original Fadeev-Popov gauge fixing
term with
\begin{equation*}
-\frac{1}{2\xi} \left(\partial_\mu A^{\mu\mathrm{a}}\right)^2 \to
-\frac{1}{2\xi} \left(\partial_\mu A^{\mu\mathrm{a}} + ig \xi
  \left(\phi^* t^\mathrm{a} \phivev - \phisvev t^\mathrm{a}
    \phi\right)\right)^2\ .
\end{equation*}
Furthermore, this shift gives rise to an additional
vector-vector-scalar coupling in the $(D\phi)^2$ term, and masses for
the $A_\mu$
\begin{equation*}
\left[ig A_\mu^\mathrm{a} t^\mathrm{a}
  \left(\phisvev+\phi^*\right)\right]_\ell \left[-ig A_\nu^\mathrm{b}
  t^\mathrm{b} \left(\phisvev+\phi^*\right)\right]_m g^{\mu\nu} =
g^2 A^{\mu\mathrm{a}} A^{\nu\mathrm{b}} g_{\mu\nu} t^\mathrm{a}
\left(\phisvev+\phi^*\right)_\ell t^\mathrm{b} \left(\phivev
  +\phi\right)_m\ .
\end{equation*}
Let us define 
\begin{equation*}
\left(t^\mathrm{a} \phivev \right)_\ell = v_\ell^\mathrm{a}\ .
\end{equation*}
Then, the Lagrangian contains the terms
\begin{equation*}
g^2 A^{\mu\mathrm{a}} A^{\nu\mathrm{b}} g_{\mu\nu} \left(
  t_{\ell m}^\mathrm{a} v_m^\mathrm{b} \phi_\ell^* + t_{\ell 
    m}^\mathrm{b} v_\ell^\mathrm{a} \phi_m + \left(t^\mathrm{a}
    t^\mathrm{b}\right)_{\ell m} \phi^*_\ell \phi_m\right)\ .
\end{equation*}
Since there are two ways to connect the gauge bosons in the
vector-vector-scalar term, we obtain the following Feynman rule
\begin{equation}
ig^2 g_{\mu\nu} t_{\ell m}^\mathrm{a} v_m^\mathrm{b} + ig^2 g_{\nu\mu} 
t_{\ell m}^\mathrm{b} v_m^\mathrm{a}\ .
\end{equation}

\subsubsection{Applications to the Standard Model}

How do these rules apply in the Standard Model?  Consider the $\phi^+
\Wminus \photon$ coupling.  Since $\photon_\mu = \particle{W^3}_\mu
\sin\thetaW + \particle{B}_\mu \cos\thetaW$, we obtain
\begin{equation*}
ig^2 g_{\mu\nu}\left(0 + \frac{v}{2} Q\right) \sin\thetaW = 
i (g \sin\thetaW) \frac{gv}{2} g_{\mu\nu} Q = 
ieM_\Wpart Q g_{\mu\nu}\ ,
\end{equation*}
where $Q$ is the charge of the $\phi$.  This result is in agreement
with \cite{cheng}.  Consider next the $\phi^+ \Wminus \Znaught$
coupling.  Since $\Znaught_\mu = \particle{W^3}_\mu \cos\thetaW -
\particle{B}_\mu \sin\thetaW$, and we obtain
\begin{multline*}
i g \frac{g}{\cos\thetaW} g_{\mu\nu} \left( 0 +
  \frac{v}{2\cos\thetaW} \left( T^3 - \sin^2 \thetaW Q\right) \right)
\cos\thetaW\\
= -ig \sin^2\thetaW g_{\mu\nu} Q \frac{gv}{2 \cos\thetaW}
= -ig M_\Znaught \sin^2\thetaW Q g_{\mu\nu}\ ,
\end{multline*}
where, again, $Q$ is the charge of the $\phi$.  This result, too, is
in agreement with \cite{cheng}.

\subsubsection{Connection between
  $\particle{V^\mu}\particle{f}\particle{f'}$ and
  $\particle{S_V}\particle{f}\particle{f'}$} 

When calculating matrix elements of interactions with internal massive
vector bosons, $\particle{V^\mu}$, we often need to add additional
diagrams with the vectors replaced by the unphysical scalars,
$\particle{S_V}$, that exist in the theory to maintain unitarity.  The
question then arises, ``Given the couplings between a pair of fermions
and a vector, say
$ig\DG{\mu}\left(a^\mathrm{a}+b^\mathrm{a}\DG{5}\right)$ for the
interaction $\particle{f_i}\to\particle{V^\mu}\particle{f_e}$, how do
we extract the corresponding coupling
$ig\left(A^\mathrm{a}+B^\mathrm{a}\DG{5}\right)$ for the interaction
$\particle{f_i}\to\particle{S_V}\particle{f_e}$?''  Well, since an
$S$-matrix (or, correspondingly, a $T$-matrix) element must be gauge
independent (independent of the chosen $R_\xi$ gauge parameter), if we
can find a set of diagrams involving the vertices of interest, we can
find the couplings $A$ and $B$ (from here on, we suppress color
indices).  In particular, consider the $t$-channel \particle{V^\mu}\ 
exchange in the process $\particle{f_i}\particle{f'_i} \to
\particle{f_e}\particle{f'_e}$.  Before writing down this matrix
element, we must express the propagators of the vector and scalar.
First the vector
\begin{align*}
D_{\mu\nu}^{\mathrm{ab}}(k) & = \frac{-i \delta^{\mathrm{ab}}}{k^2-M^2}
\left(g_{\mu\nu} -\frac{k_\mu k_\nu}{k^2-\xi M^2}(1-\xi)\right)\\
& = \frac{-i \delta^{\mathrm{ab}}}{k^2-M^2}
\left(g_{\mu\nu} -\frac{k_\mu k_\nu}{k^2}\right) +\frac{i
  \delta^{\mathrm{ab}} \xi}{k^2-\xi M^2} \left(\frac{k_\mu
    k_\nu}{k^2}\right)\\ 
& = \frac{-i \delta^{\mathrm{ab}}}{k^2-M^2}
\left(g_{\mu\nu} -\frac{k_\mu k_\nu}{k^2}\right) - \frac{i
  \delta^{\mathrm{ab}}}{k^2-\xi M^2} \left(\frac{k_\mu
    k_\nu}{M^2}\right)\ ,
\intertext{and second, the scalar}
D^{\mathrm{ab}}(k) & = \frac{i \delta^{\mathrm{ab}}}{k^2-\xi M^2}\ . 
\end{align*}

The previously mentioned matrix element, then, is given by
\begin{multline*}
iM = \left[\Spinorubar  ig \DG{\mu} \left(a + b\DG{5}\right)
  \Spinoru\right] \left[ \Spinorubark ig \DG{\nu}
  \left(c+d\DG{5}\right) \Spinoruk\right]\\
\times \left[\frac{-i}{q^2-M^2}
  \left(g_{\mu\nu} - \frac{k_\mu k_\nu}{M^2}\right) - \frac{i q_\mu
    q_\nu}{M^2} \left(\frac{1}{q^2 - \xi M^2}\right)\right]\\
+ \left[\Spinorubar ig\left(A+B\DG{5}\right) \Spinoru\right]
\left[\Spinorubark ig\left(C+D\DG{5}\right) \Spinoruk\right]
\frac{i}{q^2-\xi M^2}\ .
\end{multline*}
The $\xi$ cancellation requires the second piece of the vector
propagator and the scalar propagator to cancel.  Therefore, we must
have
\begin{equation*}
\Spinorubar ig \left(A+B\DG{5}\right) \Spinoru = \pm \Spinorubar
\frac{ig}{M} \qslash \left(a+b\DG{5}\right) \Spinoru\ .
\end{equation*}
Suppressing the spinor notation,
\begin{align*}
A+B\DG{5} & = \pm \frac{1}{M}\left(\pslash -
  \ppslash\right)\left(a+b\DG{5}\right)\\
& = \pm \frac{1}{M} \left\{\left(a-b\DG{5}\right)\pslash - \ppslash
  \left(a+b\DG{5}\right) \right\}\\
& = \pm \frac{1}{M} \left\{ \left(m_\mathrm{i} - m_\mathrm{e}\right)a
  - \left(m_\mathrm{i} + m_\mathrm{e}\right) b\DG{5} \right\}\ .
\end{align*}
The sign ambiguity is superfluous in the situation discussed here; we
need only pick one convention and stick with it.  Since these vertices
sit at the ends of internal lines, they will come in pairs and the
signs will cancel, even in matrix elements, regardless of the
convention.  However, if we need to include other goldstone vertices
in our diagrams, we must choose a globally consistent sign.  For
consistency with the sign used in the vector-vector-goldstone case
above, we must choose the minus sign in this note.

The Feynman rule for $\particle{f_i} \to \particle{f_e}
\particle{S_V}$, then, will be given by $i g \left(A + B
  \DG{5}\right)$, where
\begin{gather}
A = \frac{m_\mathrm{i} - m_\mathrm{e}}{M} a \qquad\qquad B = -
\frac{m_\mathrm{i} + m_\mathrm{e}}{M} b\ .
\end{gather}
The Feynman rule for the conjugate process, $\particle{f_e} \to
\particle{f_i} \particle{S_V}$, will be given by $-i g \left(
  A^\dagger - B^\dagger \DG{5}\right)$ (note the minus sign with the
\DG{5}).

\subsection{Kinematic Conventions}

\begin{figure}
\begin{center}
\includegraphics[width=(\textwidth/2-1in)]{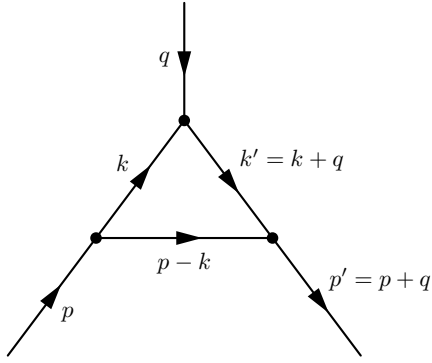}
\caption{Definition of our kinematic variable convention.}
\label{fig:momenta}
\end{center}
\end{figure}

Since we will be looking only at the one-loop, three vertex triangle
diagrams, it will simplify our task if we settle on kinematic
definitions once.  We define the momenta as in
Figure~\ref{fig:momenta}.  In particular, the incoming fermion carries
momentum $p$ into the vertex (lower left line), the outgoing fermion
carries momentum $p'$ out of the vertex (lower right line), and the
photon carries momentum $q$ into the vertex (top line).  The momenta
carried on the internal lines are also displayed in the figure; these
will be either fermions or bosons as appropriate, but we will use the
same notation for the momenta in any case.

\subsection{Feynman Parameter Reduction}

In calculating the amplitudes for the various diagrams of interest, we
always find the product of three propagator denominators (we will make
the narrow width approximation, $\Gamma = 0$).  These denominators
will need to be combined into a result which can be integrated over
the internal momentum of the loop.  We will simplify the general
expression obtained as much as possible.

We note the Feynman Parameter result for three distinct denominators 
\begin{equation}
\frac{1}{ABC} = \int\limits_0^1 \dd{x}\dd{y}\dd{z} \delta(1-x-y-z)
\frac{2}{\left(xA+yB+zC\right)^3} = \int\limits_0^1 \dd{x}\dd{y}\dd{z} 
\delta(1-x-y-z) \frac{2}{D^3}\ .
\end{equation}
In the case of two fermions and one boson in the loop, the term $ABC$
is given by
\begin{equation*}
ABC = \left(k^2-m_\internal^2\right) \left((k+q)^2-m_\internal^2\right) 
\left((p-k)^2-M^2\right)\ ,
\end{equation*}
which gives a $D$ term
\begin{align*}
D & = x\left(k^2-m_\internal^2\right)+ y\left({k'}^2-m_\internal^2\right)+ 
z\left((p-k)^2-M^2\right)\\
& = (x+y+z)k^2 -(x+y)m_\internal^2 -zM^2 +zp^2 +yq^2 + 2k(yq-zp)\\
\intertext{substituting $p^2=m_\external^2$, $x+y+z=1$, and
  completing the square, we find} 
& = \left(k+yq-zp\right)^2 - \left(yq-zp\right)^2 +yq^2
-(x+y)m_\internal^2 -zM^2 +z m_\external^2\ ,
\end{align*}
Next, expand the $-(yq-zp)^2+yq^2$ term
\begin{align*}
yq^2 - \left(yq-zp\right)^2 & = yq^2 - y^2q^2 +2yzqp -z^2p^2\\
& = xyq^2 - z^2m_\external^2\ .
\end{align*}
Thus, we find
\begin{align}
D & = \left(k+yq-zp\right)^2 +xyq^2 -z^2m_\external^2 +zm_\external^2
-zM^2 -(x+y)m_\internal^2 \nonumber \\
& = \ell^2 - \Delta +xyq^2\ ,
\end{align}
where we have
\begin{gather}
\ell = k+yq-zp
\label{eqn:defl}\\
\Delta = z(z-1) m_\external^2 + zM^2 +(x+y) m_\internal^2\ .
\end{gather}
If we define $u=x+y$, then we can simplify this expression to read 
\begin{equation}
\Delta = u(u-1) m_\external^2 +(1-u)M^2 + um_\internal^2\ .
\end{equation}

In the case of two internal boson lines and one internal fermion line, 
we simply swap $m_\internal$ for $M$, and obtain
\begin{equation}
\Delta = u(u-1) m_\external^2 +(1-u)m_\internal^2 + uM^2\ .
\end{equation}

As can be seen above, all of the integrals will be functions of $u =
x+y$ alone.  Therefore, it makes sense for us to change variables so
that we can trivially reduce the number of integrals we have to
perform.  In particular, we find
\begin{equation*}
F_2(0) = \int\limits_0^1 \dd{x} \int\limits_0^1 \dd{y}
\int\limits_0^1 \dd{z} \delta(1-x-y-z) f(x+y)
= \int\limits_0^1 \dd{x} \int\limits_0^{1-x} \dd{y} f(x+y)\ .
\end{equation*}
Now, with the previous definition $u=x+y$ and defining $v=x-y$,
\begin{equation*}
F_2(0) = \int\limits_0^1 \dd{u} \int\limits_{-u}^{u} \dd{v} f(u) |J|\ .
\end{equation*}
The Jacobian determinant, $J$, is given by
\begin{equation*}
J(u,v) = \left| \begin{array}{cc} \frac{\partial u}{\partial x}
    & \frac{\partial u}{\partial y} \\ \frac{\partial v}{\partial x} &
    \frac{\partial v}{\partial y} \end{array}\right| = -\frac{1}{2}\ .
\end{equation*}
And we find
\begin{equation}
F_2(0) = \int\limits_0^1 \dd{u} \int\limits_{-u}^{u} \dd{v} f(u)
|J|
= \int\limits_0^1 \dd{u} 2u \left|-\frac{1}{2}\right| f(u)
= \int\limits_0^1 \dd{u} u f(u)\ ,
\end{equation}
which vastly simplifies the integrals obtained in the calculations. 

\subsection{Relations Encountered in Numerator Reductions}

In the previous section we changed integration variables from $k$ to
$\ell$, defined in Equation~\ref{eqn:defl}.  We will need to use many
relations to eliminate $k$ and $k'$ in favor of $\ell$, $p$, and $p'$,
beginning with
\begin{gather}
k = \ell -yp' +(1-x)p\\
k' = \ell +(1-y)p' -xp\ ,
\end{gather}
when calculating the amplitudes that contain $F_2(0)$.  In this
section, we derive a number of more complicated expressions, in
advance of our need for them in Section~\ref{sec:building}.

We also encounter many Dirac Matrix expressions that must be reduced
to functions of $\DG{\mu}$, $p^\mu$, and ${p'}^\mu$.  Below, we derive
a number of relations that arise in our derivations
\begin{gather*}
\Spinorubar \pslash \DG{\mu} \Spinoru = \Spinorubar \left(2p^\mu
  -\DG{\mu} \pslash \right) \Spinoru = \Spinorubar \left(2p^\mu
  -m_\external \DG{\mu} \right) \Spinoru\\
\Spinorubar \DG{\mu} \ppslash \Spinoru = \Spinorubar \left(2{p'}^\mu
  -\ppslash \DG{\mu} \right) \Spinoru = \Spinorubar \left(2{p'}^\mu
  -m_\external \DG{\mu} \right) \Spinoru\ ,
\end{gather*}
where we have used the Dirac gamma matrix relation
\begin{equation*}
\anticommutator{\DG{\mu}}{\DG{\nu}} = 2 g^{\mu\nu}
\end{equation*}
We will now derive a number of similar results, suppressing the spinors.
\begin{align*}
\pslash \DG{\mu} \ppslash & = \pslash \left(2{p'}^\mu
  -\ppslash\DG{\mu}\right) = 2m_\external {p'}^\mu - \pslash \ppslash
\DG{\mu}\\
& = 2 m_\external \left(p+p'\right)^\mu - m_\external^2 - 2 p \cdot p' 
\DG{\mu}\ .
\end{align*}
But, we have
\begin{gather*}
q^2 = \left(p'-p\right)^2 = {p'}^2+p^2 -2p \cdot p' = 2m_\external^2 - 2 
p \cdot p'\\
\to -2 p \cdot p' = q^2 - 2m_\external^2\ ,
\end{gather*}
so we finally obtain
\begin{equation*}
\pslash \DG{\mu} \ppslash = 2 m_\external \left(p'+p\right)^\mu +
\left(q^2 -3m_\external^2\right) \DG{\mu}\ ,
\end{equation*}
and likewise
\begin{equation*}
\ppslash \DG{\mu} \pslash = m_\external^2 \DG{\mu}\ .
\end{equation*}
There are a number of additional results with two slashed $p$ momenta
that we will need
\begin{equation*}
\ppslash \ppslash \DG{\mu}  = m_\external^2 \DG{\mu} = \DG{\mu}
\pslash \pslash \ . 
\end{equation*}

\begin{align*}
\ppslash \pslash \DG{\mu} & = m_\external \left( 2p^\mu - m_\external
  \DG{\mu}\right) \\
& = 2 m_\external p^\mu - m_\external^2 \DG{\mu}\ .
\end{align*}

\begin{align*}
\pslash \ppslash \DG{\mu} & = \left(2 p \cdot p' - \ppslash \pslash
\right) \DG{\mu}\\
& = \left(2m_\external^2 -q^2\right) \DG{\mu} - m_\external
\left(2p^\mu -m_\external \DG{\mu}\right)\\
& = \left(3 m_\external^2 - q^2\right) \DG{\mu} - 2 m_\external p^\mu\ 
.
\end{align*}

\begin{align*}
\pslash \pslash \DG{\mu} & = \pslash  \left(2 p^\mu -m_\external
  \DG{\mu} \right)\\
& = 2m_\external p^\mu - m_\external \left(2 p^\mu -m_\external
  \DG{\mu}\right)\\
& = m_\external^2 \DG{\mu}\\
& = \DG{\mu} \ppslash \ppslash\ .
\end{align*}

\begin{align*}
\DG{\mu} \ppslash \pslash & = m_\external \left( 2 {p'}^\mu -m_\external 
\DG{\mu}\right)\\
& = 2 m_\external {p'}^\mu - m_\external^2 \DG{\mu}\ .
\end{align*}

\begin{align*}
\DG{\mu} \pslash \ppslash & = \DG{\mu} \left(2 p' \cdot p - \ppslash
  \pslash\right)\\
& = \left(2 m_\external^2 - q^2\right) \DG{\mu} - m_\external
\left(2{p'}^\mu -m_\external \DG{\mu}\right)\\
& = \left(3 m_\external^2 - q^2\right) \DG{\mu} - 2m_\external {p'}^\mu\ 
.
\end{align*}

The next relations involve single terms in $\kslash$ and $\kpslash$.
\begin{align*}
\kslash \DG{\mu} & = \left(\lslash -y\ppslash
  +(1-x)\pslash\right)\DG{\mu}\\
& = \lslash \DG{\mu} - ym_\external \DG{\mu} +(1-x)\left(2p^\mu -
  m_\external \DG{\mu}\right)\\
& = \lslash \DG{\mu} + 2(1-x)p^\mu -\left(1-(x-y)\right) m_\external
\DG{\mu}\ .
\end{align*}

\begin{align*}
\DG{\mu} \kslash & = \DG{\mu} \left(\lslash -y\ppslash
  +(1-x)\pslash\right)\\
& = \DG{\mu} \lslash -y\left(2{p'}^\mu - m_\external \DG{\mu}\right)
+(1-x) m_\external \DG{\mu}\\
& = \DG{\mu} \lslash -2y {p'}^\mu +\left(1-(x-y)\right) m_\external
\DG{\mu}\ .
\end{align*}

\begin{align*}
\kpslash \DG{\mu} & = \left(\lslash +(1-y)\ppslash -x\pslash \right)
\DG{\mu}\\
& = \lslash \DG{\mu} +(1-y) m_\external \DG{\mu} -x\left(2p^\mu 
  -m_\external \DG{\mu}\right)\\
& = \lslash \DG{\mu} - 2xp^\mu +\left(1+(x-y)\right) m_\external
\DG{\mu}\ . 
\end{align*}

\begin{align*}
\DG{\mu} \kpslash & = \DG{\mu} \left(\lslash +(1-y)\ppslash -x\pslash
\right)\\
& = \DG{\mu} \lslash +(1-y)\left(2{p'}^\mu -m_\external \DG{\mu}\right)
-xm_\external \DG{\mu}\\
& = \DG{\mu} \lslash + 2(1-y){p'}^\mu - \left(1+(x-y)\right) m_\external 
\DG{\mu}\ .
\end{align*}

These results for slashed $k$ and $k'$ terms will appear in the
amplitudes we calculate the combinations below.  Additionally, terms
in $F_2(0)$ which depend on only one factor of $\ell$ (or \lslash)
will yield zero on integration over $\dd[4]{\ell}$.  Thus, from here
on we will drop all terms which carry a single $\ell$.  When we drop
terms, we will note that we have done so by using the notation
$\Rightarrow$.  Referring back to the definition of the fermion-photon
vertex function, note that we are looking for the coefficients of
terms $\sigma^{\mu\nu} q_\nu$, Equation~\ref{eqn:definition}.  We will
obtain these by finding terms in $p'+p$, and using the Gordon Identity
\cite{peskin}
\begin{equation}
\Spinorubar \DG{\mu} \Spinoru = \frac{1}{2m_\external} \Spinorubar
\left( \left(p'+p\right)^\mu + i\sigma^{\mu\nu} q_\nu \right)
\Spinoru\ ,
\end{equation}
to exchange those for the terms we want.  The extra terms in
$\DG{\mu}$ that arise when we perform this substitution contribute to
$F_1(0)$, and can also be dropped in our calculations.

\begin{gather*}
\kslash \DG{\mu} + \DG{\mu} \kpslash \Rightarrow 2(1-x)p^\mu + 2(1-y){p'}^\mu
\Rightarrow \left(2-(x+y)\right)\left(p'+p\right)^\mu =
(2-u)\left(p'+p\right)^\mu\\
\DG{\mu}\kslash +\kpslash\DG{\mu} \Rightarrow -2y{p'}^\mu -2xp^\mu \Rightarrow
-(x+y)\left(p'+p\right)^\mu = -u \left(p'+p\right)^\mu\ .
\end{gather*}

\begin{align*}
\kslash \DG{\mu} \kpslash & = \left[\lslash -y\ppslash
  +(1-x)\pslash\right] \DG{\mu} \left[\lslash +(1-y)\ppslash
  -x\pslash\right]\\
& = \left[\lslash -ym_\external +(1-x)\pslash\right] \DG{\mu}
\left[\lslash +(1-y)\ppslash -xm_\external\right]\\
& \Rightarrow -y(1-y)m_\external \DG{\mu} \ppslash -x(1-x)m_\external \pslash 
\DG{\mu} +(1-x)(1-y) \pslash \DG{\mu} \ppslash\\
& \Rightarrow -2y(1-y)m_\external {p'}^\mu -2x(1-x)m_\external p^\mu
+2(1-x)(1-y)m_\external \left(p'+p\right)^\mu\\
& = -m_\external \left(y(1-y) +x(1-x) -2(1-x)(1-y)\right)
\left(p'+p\right)^\mu + m_\external \left(-x(1-x)
  +y(1-y)\right)\left(p'-p\right)^\mu\\
\intertext{but since the final term is odd in $x\leftrightarrow y$
  while the denominator and integrals are even, the $p'-p$ term
  vanishes, so}
& = -m_\external \left(p'+p\right)^\mu \left(y+x -y^2-x^2-2xy -2
  +2(x+y)\right)\\
& = -m_\external \left(p'+p\right)^\mu \left(2(x+y)-(x+y)^2-2\right)\\
\intertext{Performing the $u=x+y$ substitution}
& = m_\external \left(p'+p\right)^\mu \left(u^2-3u+2\right)\\
& = m_\external \left(p'+p\right)^\mu (u-1)(u-2)\ .
\end{align*}

And finally,
\begin{align*}
\kpslash \DG{\mu} \kslash & = \left[\lslash +(1-y)\ppslash
  -x\pslash\right] \DG{\mu} \left[\lslash -y\ppslash
  +(1-x)\pslash\right]\\
& \Rightarrow \left[(1-y) m_\external -x\pslash \right] \DG{\mu}
\left[-y\ppslash +(1-x)m_\external\right]\\
& \Rightarrow -y(1-y)m_\external \DG{\mu} \ppslash -x(1-x)m_\external \pslash 
\DG{\mu} +xy\pslash \DG{\mu} \ppslash \\
& \Rightarrow -2y(1-y) m_\external {p'}^\mu -2x(1-x) m_\external p^\mu + 2xy
m_\external \left(p'+p\right)^\mu\\
& = m_\external \left(-y(1-y) -x(1-x)
  +2xy\right)\left(p'+p\right)^\mu\\
& = m_\external \left(p'+p\right) \left((x+y)^2-(x+y)\right)\\
\intertext{Performing the $u=x+y$ substitution, we find}
& = m_\external \left(p'+p\right)^\mu \left(u^2-u\right)\\
& = m_\external \left(p'+p\right)^\mu u(u-1)\ .
\end{align*}

To summarize, then, the contributions to the numerator of $F_2(0)$ arising
from each of these terms are
\begin{gather}
\DG{\mu} \pslash \Mapsto 0\\
\DG{\mu} \ppslash \Mapsto 2 {p'}^\mu\\
\pslash \DG{\mu} \Mapsto 2 p^\mu\\
\ppslash \DG{\mu} \Mapsto 0\\
\ppslash \ppslash \DG{\mu} \Mapsto 0\\
\ppslash \pslash \DG{\mu} \Mapsto 2 m_\external p^\mu\\
\pslash \ppslash \DG{\mu} \Mapsto - 2m_\external p^\mu\\
\pslash \pslash \DG{\mu} \Mapsto 0\\
\DG{\mu} \pslash \pslash \Mapsto 0\\
\DG{\mu} \ppslash \pslash \Mapsto 2 m_\external {p'}^\mu\\
\DG{\mu} \pslash \ppslash \Mapsto -2 m_\external {p'}^\mu\\
\DG{\mu} \ppslash \ppslash \Mapsto 0\\
\kslash \DG{\mu} + \DG{\mu} \kpslash \Mapsto -i \sigma^{\mu\nu} q_\nu
(2-u) = i \sigma^{\mu\nu} q_\nu \left[u-2\right]\\
\DG{\mu} \kslash +\kpslash \DG{\mu} \Mapsto i \sigma^{\mu\nu} q_\nu
\left[ u \right]\\
\kslash \DG{\mu} \kpslash \Mapsto -i \sigma^{\mu\nu} q_\nu m_\external
(u-1)(u-2) = i\sigma^{\mu\nu} q_\nu \left[-m_\external
  (u-1)(u-2)\right]\\
\kpslash \DG{\mu} \kslash \Mapsto -i\sigma^{\mu\nu} q_\nu m_\external u
(u-1) = i \sigma^{\mu\nu} q_\nu \left[ m_\external u (1-u) \right]\ . 
\end{gather}

\section{A Systematic Approach: Using the Tools}
\label{sec:building}

Having derived needed tools, or ``building blocks'' for our one loop
calculation, we now turn to the calculation of the four classes of
diagrams to \amuon: neutral and charged scalars and neutral and
charged vectors.

\subsection{Neutral/Charged Scalar Contributions}
\label{sec:scalar:neutral}

\subsubsection{Summary}

\begin{figure}
\begin{center}
\includegraphics[width=(\textwidth-1in)/2]{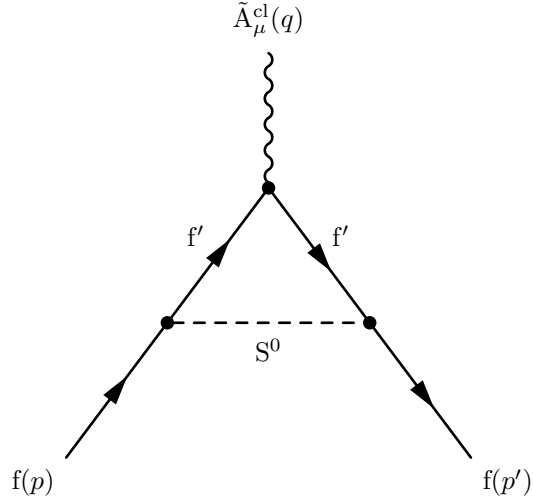}
\caption{This Feynman diagram gives a one-loop
  contribution to \afermion\ from neutral/charged scalars.}
\label{fig:scalar:neutral}
\end{center}
\end{figure}

Electrically neutral and charged scalar contributes the following term
to the anomalous magnetic moment of a charged fermion:
\begin{equation}
\aext = \frac{m_\external g^2}{8\pi^2}
\frac{Q_\internal}{Q_\external} \int\limits_0^1 \dd{u} u^2
\frac{m_\internal \left(vv^\dagger-aa^\dagger\right) - m_\external
  \left(vv^\dagger+aa^\dagger\right)(u-1) }{(1-u)M^2 +um_\internal^2
  +u(u-1)m_\external^2}\ .
\end{equation}
The Feynman diagram contributing to this term is found in
Figure~\ref{fig:scalar:neutral}. 

\subsubsection{The Calculation}

We calculate the contribution of a scalar that couples to a pair of
fermions $\particle{f_\internal}\to \particle{f_\external}
\particle{S}$ with Feyman Rule $ig(v+a\DG{5})$, that is, with
arbitrary vector and axial couplings.  The matrix element is given by
\begin{align*}
  i \amplitude & = ie Q_e \Spinorubar \Gamma^\mu \Spinoru\\
  & = \int \frac{\dd[4]{k}}{(2\pi)^4} \Spinorubar
  ig\left(v+a\DG{5}\right)
  \frac{i(\kpslash+m_\internal)}{{k'}^2-m_\internal^2} ieQ_\internal
  \DG{\mu} \frac{i(\kslash+m_\internal)}{k^2-m_\internal^2}
  ig\left(v^\dagger-a^\dagger\DG{5}\right) \Spinoru
  \frac{i}{(p-k)^2-M^2}\\
  & = ieQ_\external \frac{Q_\internal}{Q_\external} \int
  \frac{\dd[4]{k}}{(2\pi)^4} g^2 i^5 \Spinorubar
  \frac{\left(v+a\DG{5}\right) \left(\kpslash+m_\internal\right)
    \DG{\mu} \left(\kslash+m_\internal\right)
    \left(v^\dagger-a^\dagger\DG{5}\right)}{\left(k^2-m_\internal^2\right)
    \left({k'}^2-m_\internal^2\right) \left((p-k)^2-M^2\right)}
  \Spinoru\ .
\end{align*}
Searching now for the $p^\mu$ and ${p'}^\mu$ terms (suppressing the
spinors), and dropping irrelevant terms quickly, we find
\begin{multline*}
\left(v+a\DG{5}\right) \left(\kpslash+m_\internal\right)
    \DG{\mu} \left(\kslash+m_\internal\right)
    \left(v^\dagger-a^\dagger\DG{5}\right) =\\ vv^\dagger
    \left(\kpslash+m_\internal\right) \DG{\mu}
    \left(\kslash+m_\internal\right)   + aa^\dagger
    \left(\kpslash-m_\internal\right) \DG{\mu}
    \left(\kslash-m_\internal\right)\ .
\end{multline*}
Thus, we need to find
\begin{equation*}
\kpslash \DG{\mu} \kslash \pm m_\internal\left(\kpslash \DG{\mu}+
  \DG{\mu} \kslash\right)\ .
\end{equation*}
Using the results from earlier sections of this note, we quickly find
the result
\begin{equation*}
\frac{i \sigma^{\mu\nu} q_\nu}{2 m_\external} 2 m_\external \left[
  -m_\external u (u-1) \pm m_\internal u \right]\ .
\end{equation*}
Thus, the relevant remaining numerator pieces result in
\begin{equation*}
\Spinorubar \left\{ \cdots \right\} \Spinoru \Mapsto \Spinorubar
\frac{i \sigma^{\mu\nu} q_\nu}{2 m_\external} 2 m_\external
\left[-m_\external u (u-1)\left(vv^\dagger+aa^\dagger\right) +
  m_\internal u \left(vv^\dagger-aa^\dagger\right) \right]\ .
\end{equation*}

We have almost arrived at our final result
\begin{multline*}
i \amplitude \supset ieQ_\external \Spinorubar \frac{i \sigma^{\mu\nu}
  q_\nu}{2 m_\external} \left\{ \int\frac{\dd[4]{\ell}}{(2\pi)^4}
  \int\limits_0^1 \dd{u} ig^2 u \frac{Q_\internal}{Q_\external}
  \frac{2}{\left(\ell^2-\Delta+xyq^2\right)^3}\right.\\
  \left. 2m_\external \left[ 
    -m_\external u (u-1)\left(vv^\dagger+aa^\dagger\right) +
    m_\internal u \left(vv^\dagger-aa^\dagger\right)  \right]
\right\} \Spinoru\ ,
\end{multline*}
from which we can extract the $g-2$ contribution, since
\begin{equation*}
i \amplitude \supset ieQ_\external \Spinorubar \frac{i \sigma^{\mu\nu}
  q_\nu}{2 m_\external} F_2(q^2) \Spinoru\ .
\end{equation*}
Performing the $\dd[4]{\ell}$ integral, and taking the $q^2\to 0$
limit, we find our final result
\begin{equation}
\aext = F_2(0) = \frac{m_\external g^2}{8\pi^2}
\frac{Q_\internal}{Q_\external} \int\limits_0^1 \dd{u} u^2
\frac{m_\internal \left(vv^\dagger-aa^\dagger\right) - m_\external
  \left(vv^\dagger+aa^\dagger\right)(u-1) }{(1-u)M^2 +um_\internal^2
  +u(u-1)m_\external^2}\ .
\end{equation}

\subsection{Charged Scalar Contributions}
\label{sec:scalar:charged}

\subsubsection{Summary}

\begin{figure}
\begin{center}
\includegraphics[width=(\textwidth-1in)/2]{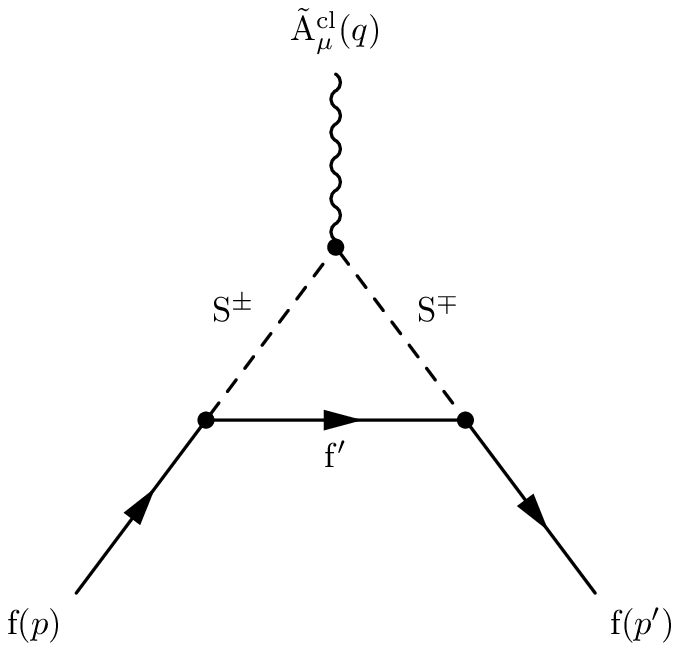}
\caption{This Feynman diagram gives a one-loop contribution to
  \afermion\ from charged scalars.}
\label{fig:scalar:charged}
\end{center}
\end{figure}

An electrically charged scalar contributes the following term to the
anomalous magnetic moment of a charged fermion:
\begin{equation}
\aext = \frac{-m_\external g^2}{8\pi^2}
\frac{Q_\particle{S}}{Q_\external} \int\limits_0^1 \dd{u} u(1-u)
\frac{\left(vv^\dagger+aa^\dagger\right) u m_\external +
  \left(vv^\dagger-aa^\dagger\right) m_\internal}{uM^2
  +u(u-1)m_\external^2 +(1-u)m_\internal^2}\ .
\end{equation}
The Feynman diagram giving rise to this contribution is shown in
Figure~\ref{fig:scalar:charged}. 

\subsubsection{The Calculation}

We calculate here the contributions of an electrically charged scalar
that couples to a pair of fermions $\particle{f_\internal} \to
\particle{f_\external}\particle{S}$ with Feynman Rule $ig(v+a\DG{5})$.
The matrix element is given by
\begin{align*}
i \amplitude & = ieQ_\external \Spinorubar \Gamma^\mu \Spinoru\\
& = \int \frac{\dd[4]{k}}{(2\pi)^4} \Spinorubar
ig\left(v+a\DG{5}\right) \frac{i \left(\pslash-\kslash
    +m_\internal\right)}{(p-k)^2-m_\internal^2}
ig\left(v^\dagger-a^\dagger\DG{5}\right) \Spinoru \frac{i}{{k'}^2-M^2}
\frac{i}{k^2-M^2} ieQ_\particle{S} \left(k+k'\right)^\mu\\
& = ieQ_\external \frac{Q_\particle{S}}{Q_\external} g^2i^5
\int\frac{\dd[4]{k}}{(2\pi)^4} \Spinorubar
\frac{\left(v+a\DG{5}\right) \left(\pslash-\kslash +m_\internal\right) 
  \left(v^\dagger-a^\dagger\DG{5}\right)
  \left(k+k'\right)^\mu}{\left({k'}^2-M^2\right) \left(k^2-M^2\right)
  \left((p-k)^2-m_\internal^2\right)} \Spinoru\ .
\end{align*}
Again, we search for $p^\mu$ and ${p'}^\mu$ terms (suppressing spinors), 
and dropping irrelevant terms quickly, we find
\begin{equation*}
\pm\left(\pslash -\kslash\right) +m_\internal \Rightarrow \pm u m_\external + 
m_\internal\ .
\end{equation*}
Combining the relevant numerator pieces, we arrive at
\begin{equation*}
\Spinorubar \left\{ \cdots \right\} \Spinoru \Mapsto \Spinorubar
\frac{-i \sigma^{\mu\nu} q_\nu}{2 m_\external} \left\{
  \left(vv^\dagger+aa^\dagger \right) u m_\external
  +\left(vv^\dagger-aa^\dagger\right) m_\internal \right\}(1-u) 2
m_\external \Spinoru\ .
\end{equation*}

We now reconstruct our intermediate result
\begin{multline*}
i \amplitude \supset ie Q_\external \Spinorubar \frac{i
  \sigma^{\mu\nu} q_\nu}{2 m_\external} \left\{ \int
  \frac{\dd[4]{\ell}}{(2\pi)^4} \int\limits_0^1 \dd{u} u \left(-
    \frac{Q_\particle{S}}{Q_\external}\right) ig^2
  \frac{2}{\left(\ell^2-\Delta+xyq^2\right)}\right. \\
\left. 2 m_\external (1-u) \left[\left(vv^\dagger+aa^\dagger\right) u
    m_\external + \left(vv^\dagger-aa^\dagger\right) m_\internal
  \right]\right\} \Spinoru\ .
\end{multline*}
Performing the $\dd[4]{\ell}$ integral and taking the $q^2\to0$ limit, 
we obtain our final result
\begin{equation}
\aext = F_2(0) = \frac{-m_\external g^2}{8\pi^2}
\frac{Q_\particle{S}}{Q_\external} \int\limits_0^1 \dd{u} u(1-u)
\frac{\left(vv^\dagger+aa^\dagger\right) u m_\external +
  \left(vv^\dagger-aa^\dagger\right) m_\internal}{uM^2
  +u(u-1)m_\external^2 +(1-u)m_\internal^2}\ .
\end{equation}

\subsection{\Znaught-like Contributions}
\label{sec:vector:zlike}

\subsubsection{Summary}

\begin{figure}
\begin{center}
\begin{minipage}{\textwidth}
\subfigure[Physical Vector
Contribution]{\includegraphics[width=(\textwidth-1in)/2]{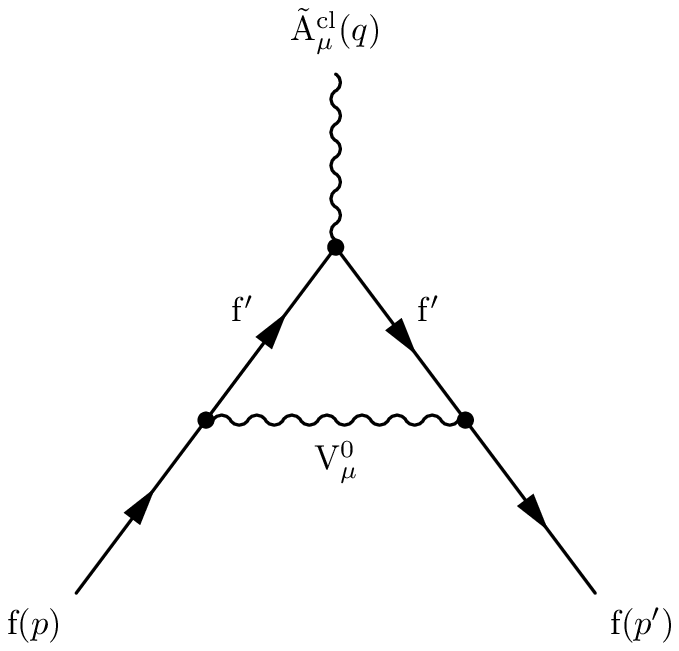}
  \label{fig:vector:zlike:vector}} \qquad\qquad 
\subfigure[Unphysical Scalar
Contribution]{\includegraphics[width=(\textwidth-1in)/2]{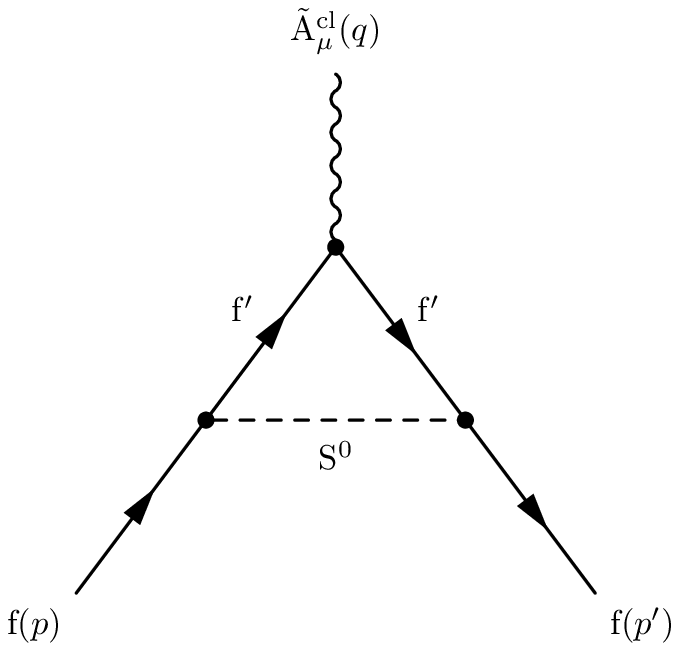}
  \label{fig:vector:zlike:scalar}} 
\end{minipage}
\caption{The diagrams for one-loop neutral and charged vector
  contributions to \afermion.  The left diagram is the contribution
  from the physical vector, while the right diagram is the
  contribution from the associated unphysical scalar (the goldstone
  mode).}
\label{fig:vector:zlike}
\end{center}
\end{figure}

A neutral or charged vector contributes the following term to the
anomalous magnetic moment of a charged fermion
\begin{multline}
\aext = -\frac{m_\external g^2}{8\pi^2}
\frac{Q_\internal}{Q_\external} \int\limits_0^1 \dd{u} \frac{u(u-1)
  \left[ 2m_\external (u-2)\left(vv^\dagger+aa^\dagger\right) +
    4m_\internal \left(vv^\dagger-aa^\dagger\right) \right]}{(1-u)M^2
  +um_\internal^2 +u(u-1)m_\external^2}\\
+ \frac{m_\external g^2}{8\pi^2} \frac{m_\internal}{M^2}
\frac{Q_\internal}{Q_\external} \int\limits_0^1 \dd{u} u^2
\frac{\left(m_\internal- m_\external\right)^2 vv^\dagger
  -\left(m_\internal+ m_\external\right)^2 aa^\dagger}{(1-u)M^2
  +um_\internal^2 +u(u-1)m_\external^2}\\ 
- \frac{m_\external g^2}{8\pi^2} \frac{m_\external}{M^2}
\frac{Q_\internal}{Q_\external} \int\limits_0^1 \dd{u} u^2
\frac{\left\{\left(m_\internal- m_\external\right)^2 vv^\dagger 
  +\left(m_\internal+ m_\external\right)^2
  aa^\dagger\right\}(u-1)}{(1-u)M^2 +um_\internal^2
+u(u-1)m_\external^2}\ ,
\end{multline}
where the second and third terms are included only when $M\neq 0$.
The diagrams which give this contribution are shown in
Figure~\ref{fig:vector:zlike}.

\subsubsection{The \Znaught-like Vector Contribution}

We calculate here the contribution of a vector to the anomalous
magnetic moment of a charged fermion; the Feynman diagram for this
amplitude is shown in Figure~\ref{fig:vector:zlike:vector}.  We call
this contribution \Znaught-like, since this is the contributing
diagram for \Znaught\ contribution to $g_\muon-2$.
\begin{align*}
  i \amplitude & = ieQ_\external \Spinorubar \Gamma^\mu \Spinoru\\
  & = \int \frac{\dd[4]{k}}{(2\pi)^4} \Spinorubar ig \DG{\rho}
  \left(v+a\DG{5}\right) \frac{i\left(\kpslash
      +m_\internal\right)}{{k'}^2 -m_\internal^2} ie Q_\internal
  \DG{\mu} \frac{i\left(\kslash +m_\internal\right)}{k^2
    -m_\internal^2} ig \DG{\sigma}
  \left(v^\dagger+a^\dagger\DG{5}\right) \Spinoru
  \frac{-i g_{\rho\sigma}}{(p-k)^2-M^2}\\
  & = ie Q_\external \left(\frac{Q_\internal}{Q_\external}\right) (-1)
  i^5 g^2 \Spinorubar \frac{N^\mu}{D'} \Spinoru\ ,
\end{align*}
where 
\begin{equation*}
D' = \left({k'}^2-m_\internal^2\right) \left(k^2-m_\internal^2\right)
\left((p-k)^2-M^2\right)\ ,
\end{equation*}
and
\begin{align*}
N^\mu & = \DG{\rho} \left(v+a\DG{5}\right) \left(\kpslash 
    +m_\internal\right) \DG{\mu} \left(\kslash +m_\internal\right) 
    \DG{\sigma}\left(v^\dagger+a^\dagger\DG{5}\right) g_{\rho\sigma}\\
& \equiv vv^\dagger n_+^\mu + aa^\dagger n_-^\mu\ .
\end{align*}
We now reduce the numerator terms
\begin{align*}
n_\pm^\mu & = \DG{\rho} \left(\kpslash \pm m_\internal\right) \DG{\mu} 
\left(\kslash \pm m_\internal\right) \DGL{\rho}\\
& = \DG{\rho} \kpslash \DG{\mu} \kslash \DGL{\rho} \pm m_\internal
\DG{\rho} \left( \kpslash \DG{\mu} + \DG{\mu} \kslash \right)
\DGL{\rho} + m_\internal^2 \DG{\rho} \DG{\mu} \DGL{\rho}\\
& \Rightarrow -2 \kslash \DG{\mu} \kpslash \pm 4 m_\internal
\left(k'+k\right)^\mu\\
& \Rightarrow -2 \left( -i \sigma^{\mu\nu} q_\nu m_\external (u-1)(u-2)
\right) \pm 4 m_\internal (1-u) \left(-i \sigma^{\mu\nu}
  q_\nu\right)\\
& = i \sigma^{\mu\nu} q_\nu \left[2 m_\external (u-1)(u-2) \pm 4
  m_\internal (u-1)\right]\ ,
\end{align*}
Thus, the numerator can be reduced to 
\begin{equation*}
\Spinorubar N^\mu \Spinoru = \Spinorubar \frac{i \sigma^{\mu\nu}
  q_\nu}{2 m_\external} 2 m_\external \left[ 2 m_\external (u-1)(u-2)
  \left(vv^\dagger+aa^\dagger\right) + 4 m_\internal (u-1)
  \left(vv^\dagger-aa^\dagger\right) \right] \Spinoru\ .
\end{equation*}

We now reconstruct the intermediate result
\begin{multline*}
i \amplitude \supset ie Q_\external \Spinorubar \frac{i
  \sigma^{\mu\nu} q_\nu}{2 m_\external} \Bigg\{ \int
  \frac{\dd[4]{\ell}}{(2\pi)^4} \int\limits_0^1 \dd{u} u
  \frac{Q_\internal}{Q_\external} \left(-2im_\external g^2\right)
  \frac{2}{\left(\ell^2 - \Delta +xyq^2\right)^3}\\ 
\times \left[2 m_\external (u-1)(u-2)
    \left(vv^\dagger+aa^\dagger\right) + 4m_\internal (u-1)
    \left(vv^\dagger-aa^\dagger\right)\right] \Bigg\} \Spinoru\ ,
\end{multline*}
Performing the momentum integral and taking the $q^2\to0$ limit, we
obtain the final result for this diagram
\begin{equation}
F_2(0) = -\frac{m_\external g^2}{8\pi^2}
\frac{Q_\internal}{Q_\external} \int\limits_0^1 \dd{u} \frac{u(u-1)
  \left[ 2m_\external (u-2)\left(vv^\dagger+aa^\dagger\right) +
    4m_\internal \left(vv^\dagger-aa^\dagger\right) \right]}{(1-u)M^2
  +um_\internal^2 +u(u-1)m_\external^2}\ .
\end{equation}

\subsubsection{The Unphysical Scalar Contribution}

For the case $M\neq0$, we must also add the results of
Section~\ref{sec:scalar:neutral}, with $v\to V$ and $a\to A$.  We
obtain
\begin{multline}
F_2(0) = \frac{m_\external g^2}{8\pi^2} \frac{m_\internal}{M^2}
\frac{Q_\internal}{Q_\external} \int\limits_0^1 \dd{u} u^2
\frac{\left(m_\internal- m_\external\right)^2 vv^\dagger
  -\left(m_\internal+ m_\external\right)^2 aa^\dagger}{(1-u)M^2
  +um_\internal^2 +u(u-1)m_\external^2} \\ 
- \frac{m_\external g^2}{8\pi^2} \frac{m_\external}{M^2}
\frac{Q_\internal}{Q_\external} \int\limits_0^1 \dd{u} u^2
\frac{\left\{\left(m_\internal- m_\external\right)^2 vv^\dagger 
  +\left(m_\internal+ m_\external\right)^2
  aa^\dagger\right\}(u-1)}{(1-u)M^2 +um_\internal^2
+u(u-1)m_\external^2}\ .
\end{multline}
This result is suppressed relative to the vector term when the
internal and external fermion masses are small compared to the
vector mass, and can be ignored in that limit.

\subsection{\Wpart-like Contributions}
\label{sec:vector:wlike}

\subsubsection{Summary}

\begin{figure}
\begin{center}
\begin{minipage}{\textwidth}
\begin{minipage}{\textwidth}
\subfigure[Double Vector
Contribution]{\includegraphics[width=(\textwidth-1in)/2]{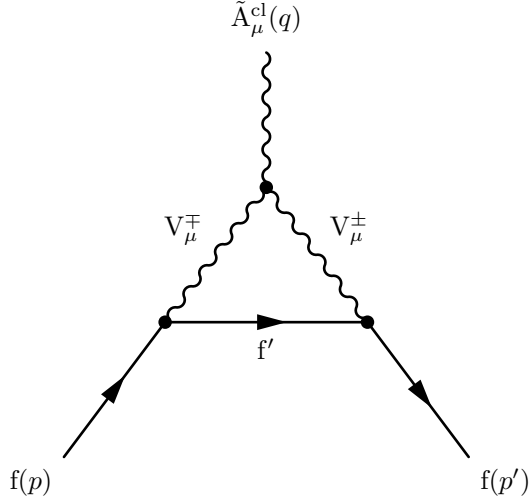}
  \label{fig:vector:wlike:vv}} 
\qquad \qquad
\subfigure[First Vector-Scalar
Contribution]{\includegraphics[width=(\textwidth-1in)/2]{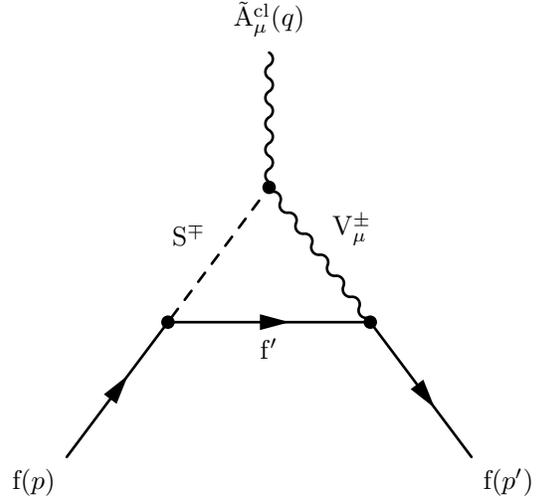}
  \label{fig:vector:wlike:vs}} 
\end{minipage}
\begin{minipage}{\textwidth}
\subfigure[Second Vector-Scalar
Contribution]{\includegraphics[width=(\textwidth-1in)/2]{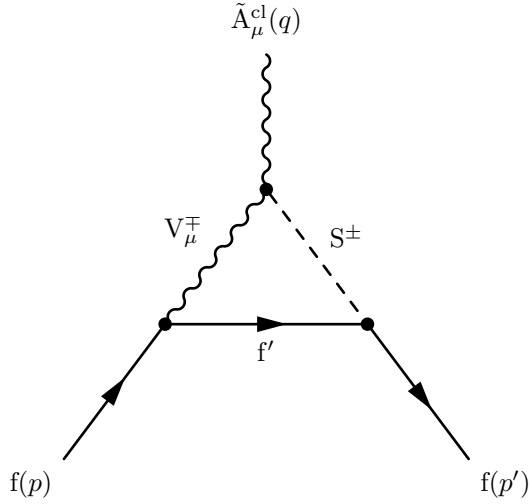}
  \label{fig:vector:wlike:sv}} 
\qquad \qquad
\subfigure[Double Scalar
Contribution]{\includegraphics[width=(\textwidth-1in)/2]{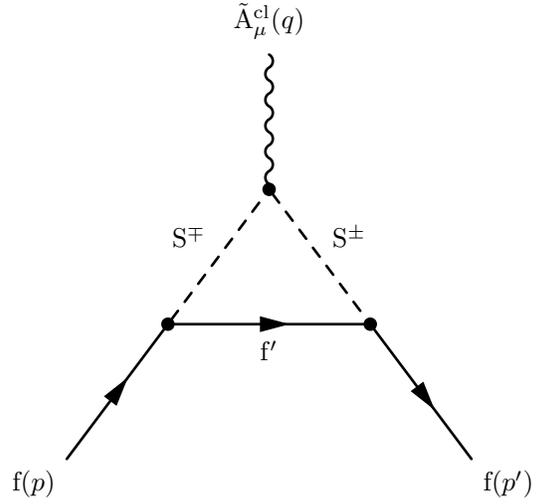}
  \label{fig:vector:wlike:ss}} 
\end{minipage}
\end{minipage}
\caption{The diagrams for one-loop charged vector contributions to
  \afermion.  The upper left diagram is the contribution from two
  physical vectors.  The upper right and lower left diagrams are the
  contributions from one physical vector and one unphysical scalar
  (the goldstone modes). The final diagram is the contribution from
  two unphysical scalars.}
\label{fig:vector:wlike}
\end{center}
\end{figure}

An electrically charged vector boson makes the following contribution
to the anomalous magnetic moment of a charged fermion
\begin{multline}
\aext = \frac{i m_\external g^2 g' f^{\mathrm{mnr}}}{4 \pi^2 e
  Q_\external} \int\limits_0^1 \dd{u} u^2 \frac{m_\external (2u+1)
  \left(vv^\dagger+aa^\dagger\right) - 3m_\internal
  \left(vv^\dagger-aa^\dagger\right)}{(1-u)m_\internal^2 + uM^2 +
  u(u-1) m_\external^2}\\
- \frac{m_\external}{8\pi^2} \frac{Q_V}{Q_\external}
\frac{{g'}^2 v^\mathrm{b} R}{eM} \int\limits_0^1 \dd{u}
\frac{u^2\left((m_\internal-m_\external) vv^\dagger
    -(m_\internal+m_\external) aa^\dagger\right) }{uM^2
  +(1-u)m_\internal^2 +u(u-1)m_\external^2}\\
-\frac{m_\external g^2}{8\pi^2} \frac{Q_V}{Q_\external} \frac{1}{M^2}
\int\limits_0^1 \dd{u} u(1-u) \left\{\frac{m_\external\left\{
    (m_\internal-m_\external)^2 vv^\dagger
    +(m_\internal+m_\external)^2 aa^\dagger \right\}
  u}{uM^2+(1-u)m_\internal^2 +u(u-1)m_\external^2}\ + \right. \\ 
\left. \frac{m_\internal \left\{(m_\internal-m_\external)^2 vv^\dagger
      - (m_\internal+m_\external)^2
      aa^\dagger\right\}}{uM^2+(1-u)m_\internal^2
    +u(u-1)m_\external^2} \right\}
\ ,
\end{multline}
where the first line is the vector-vector term, the second line
contains the contributions of the two vector-scalar terms, and the
final two lines are the contribution from the scalar-scalar term; the
terms containing scalar contributions are only included when the
vector mass is nonzero, $M \neq 0$.  The Feynman diagrams giving rise
to these contributions are shown in Figure~\ref{fig:vector:wlike}

\subsubsection{The Vector-Vector Contribution}

In this subsection, we calculate the two vector contribution to the
anomalous magnetic moment of a charged fermion.  The Feynman diagram
for this amplitude is given in Figure~\ref{fig:vector:wlike:vv}.
\begin{multline*}
i \amplitude = \int \frac{\dd[4]{k}}{(2\pi)^4} \Spinorubar
ig\DG{\alpha} \left(v+a\DG{5}\right) \frac{i\left(\pslash -\kslash
    +m_\internal\right)}{(p-k)^2-m_\internal^2} ig \DG{\beta}
\left(v^\dagger+a^\dagger\DG{5}\right) \Spinoru \\
\times \frac{-ig_{\alpha\rho}}{{k'}^2-M^2} \frac{-ig_{\beta\nu}}{k^2-M^2}
g'f^{\mathrm{mnr}} \left[g^{\mu\nu} (q-k)^\rho +g^{\nu\rho}(k+k')^\mu
  +g^{\rho\mu}(-k'-q)^\nu\right] \equiv \\
ieQ_\external \frac{1}{ieQ_\external} i^5 g^2 g' f^{\mathrm{mnr}} \int 
\frac{\dd[4]{k}}{(2\pi)^4} \Spinorubar \frac{N^\mu}{D'} \Spinoru\ ,
\end{multline*}
where
\begin{equation*}
D' = \left({k'}^2-M^2\right) \left(k^2-M^2\right)
\left((p-k)^2-m_\internal^2\right)
\end{equation*}
and
\begin{align*}
N^\mu & = \DGL{\rho} \left(v+a\DG{5}\right) \left(\pslash -\kslash
  +m_\internal\right) \DGL{\nu} \left(v^\dagger+a^\dagger\DG{5}\right)
F^{\mu\nu\rho}\\
& = vv^\dagger \DGL{\rho} \left(\pslash -\kslash +m_\internal\right)
\DGL{\nu} F^{\mu\nu\rho} + aa^\dagger \DGL{\rho}
\left(\pslash -\kslash -m_\internal\right) \DGL{\nu}
F^{\mu\nu\rho}\\
& = vv^\dagger n_{\rho\nu}^+ F^{\mu\nu\rho} +
aa^\dagger n_{\rho\nu}^- F^{\mu\nu\rho}\ ,
\end{align*}
where we have defined
\begin{equation*}
F^{\mu\nu\rho} \equiv \left[g^{\mu\nu} (q-k)^\rho +g^{\nu\rho}(k+k')^\mu
  +g^{\rho\mu}(-k'-q)^\nu\right] \equiv 1^{\mu\nu\rho} +
2^{\mu\nu\rho} + 3^{\mu\nu\rho}
\end{equation*}
We will also need the following reductions
\begin{gather}
\pslash - \kslash \Rightarrow \pslash - \left(-y\ppslash +(1-x)\pslash\right)
  = y \ppslash + x\pslash\\
q-k \Rightarrow p'-p +yp'-(1-x)p = (1+y)p' +(x-2)\\
-k'-q \Rightarrow - \left((1-y)p' -xp +p'-p\right) = (y-2)p' +(1+x)p\\
k+k' \Rightarrow \left(1-2y\right) p' + \left(1-2x\right) p\ .
\end{gather}
We can now begin to reduce the numerator, which we will do a piece at
a time.  We will drop all terms that do not end up affecting $F_2(0)$.
\begin{align*}
n_{\rho\nu}^\pm 2^{\mu\nu\rho} & \Rightarrow \DGL{\rho} \left(y\ppslash
  +x\pslash \pm m_\internal\right) \DGL{\nu} g^{\nu\rho} (k+k')^\mu\\
& = \DG{\nu} \left(y\ppslash +x\pslash \pm m_\internal\right)
\DGL{\nu} (k+k')^\mu\\
& \Rightarrow \left[-2\left(y\ppslash +s\pslash\right) \pm
  4m_\internal\right] \left[(1-2y)p'+(1-2x)p\right]^\mu\\
& \Rightarrow \left[-2m_\external (x+y) \pm4m_\internal\right] (1-x-y)
(p'+p)^\mu\\
& = \left[-2m_\external u \pm 4m_\internal\right] (1-u) (p'+p)^\mu\ .\\
\intertext{The manipulations for the other terms are similar, and we
  only quote the final results}
n_{\rho\nu}^\pm 1^{\mu\nu\rho} 
& \Rightarrow 2 \left[m_\external (x+2y) \pm m_\internal (x-2)\right] p^\mu\\
n_{\rho\nu}^\pm 3^{\mu\nu\rho} 
& \Rightarrow 2 \left[m_\external (y+2x) \pm m_\internal (y-2)\right] {p'}^\mu\ 
.
\end{align*}

Having calculated these three pieces, we can now combine them.  First, 
combine $1^{\mu\nu\rho}$ and $3^{\mu\nu\rho}$
\begin{align*}
n_{\rho\nu}^\pm \left[ 1^{\mu\nu\rho} + 3^{\mu\nu\rho}\right] & \Rightarrow
\left\{m_\external \left[(y+2x)+(x+2y)\right] \pm
  m_\internal\left[(x-2)+(y-2)\right] \right\} (p'+p)^\mu\\
& = \left[ 3 m_\external (x+y) \pm m_\internal (x+y-4) \right]
(p'+p)^\mu\\
& = \left[ 3 m_\external u \pm m_\internal (u-4)\right] (p'+p)^\mu\ .
\end{align*}
Finally, add $3^{\mu\nu\rho}$
\begin{align*}
n_{\rho\nu}^\pm \left[\cdots \right]^{\mu\nu\rho} & \Rightarrow \left[ 3
  m_\external u \pm m_\internal (u-4) - 2 m_\external u (1-u) \pm 4
  m_\internal (1-u) \right] (p'+p)^\mu\\
& = \left[m_\external (u+2u^2) \pm m_\internal (-3u)\right]
(p'+p)^\mu\\ 
& = u \left(m_\external (2u+1) \mp 3 m_\internal\right) (p'+p)^\mu\\ 
& \Rightarrow -u \left(m_\external (2u+1) \mp 3 m_\internal\right) i
\sigma^{\mu\nu} q_\nu\ .
\end{align*}

We can now substitute these results into the amplitude
\begin{multline*}
i \amplitude \supset ieQ_\external \Spinorubar \frac{i \sigma^{\mu\nu} 
  q_\nu}{2 m_\external} \left\{ \int\frac{\dd[4]{\ell}}{(2\pi)^4}
  \int\limits_0^1 \dd{u} u \frac{-2m_\external g^2 g'
    f^{\mathrm{mnr}}}{e Q_\external} u  \frac{2}{\left( \ell^2 -\Delta 
      +xyq^2\right)^3} \times \right.\\  
\left. \left[ m_\external (2u+1)\left(vv^\dagger+aa^\dagger\right) - 3 
  m_\internal \left(vv^\dagger-aa^\dagger\right)\right] \right\}
\Spinoru\ .
\end{multline*}
From which we obatin the final result, after performing the momentum
integral and taking the $q^2\to 0$ limit
\begin{equation}
F_2(0) = \frac{i m_\external g^2 g' f^{\mathrm{mnr}}}{4 \pi^2 e
  Q_\external} \int\limits_0^1 \dd{u} u^2 \frac{m_\external (2u+1)
  \left(vv^\dagger+aa^\dagger\right) - 3m_\internal
  \left(vv^\dagger-aa^\dagger\right)}{(1-u)m_\internal^2 + uM^2 +
  u(u-1) m_\external^2}\ .
\end{equation}

\subsubsection{The Vector-Scalar Contributions}

In this subsection, we calculate the two vector-scalar contributions
to the anomalous magnetic moment of a charged fermion.  The two
diagrams that give rise to this contribution are shown in
Figures~\ref{fig:vector:wlike:vs} and~\ref{fig:vector:wlike:sv}.
\begin{multline*}
i \amplitude = -\int \frac{\dd[4]{k}}{(2\pi)^4} \Spinorubar ig
\DG{\beta} \left(v+a\DG{5}\right) \frac{i \left(\pslash -\kslash
    +m_\internal\right)}{(p-k)^2- m_\internal^2} ig \left(V^\dagger-
  A^\dagger\DG{5}\right) \Spinoru \times \\
\frac{i}{k^2-M^2} \frac{-ig_{\alpha\beta}}{{k'}^2-M^2} i{g'}^2
v^\mathrm{b} R Q_V g_{\mu\alpha}\\
- \int \frac{\dd[4]{k}}{(2\pi)^4} \Spinorubar ig
\left(V+A\DG{5}\right) \frac{i \left(\pslash -\kslash
    +m_\internal\right)}{(p-k)^2- m_\internal^2} ig \DG{\beta} \left(v+
  a \DG{5}\right) \Spinoru \times \\
\frac{i}{k^2-M^2} \frac{-ig_{\alpha\beta}}{{k'}^2-M^2} i{g'}^2
v^\mathrm{b} R Q_V g_{\mu\alpha}\ ,
\end{multline*}
where the terms above correspond to the various couplings, VEVs,  and
mixing angles in the electroweak-like sector.
\begin{equation*}
= - ie Q_\external \frac{-i^6 Q_V g^2 \left({g'}^2 R
    v^{\textrm{b}}\right)}{ie Q_\external}
\int\frac{\dd[4]{k}}{(2\pi)^4} \Spinorubar \frac{N^\mu}{D} \Spinoru\ ,
\end{equation*}
where
\begin{equation*}
D = \left((p-k)^2- m_\internal^2\right)\left(k^2-M^2\right)
\left({k'}^2-M^2\right)\ ,
\end{equation*}
and 
\begin{equation*}
N^\mu = \DG{\mu}\left(v+a\DG{5}\right) \left(\pslash-\kslash
  +m_\internal\right) \left(V^\dagger-A^\ddag\DG{5}\right) +
\left(V+A\DG{5}\right) \left(\pslash-\kslash +m_\internal\right)
\DG{\mu} \left(v^\dagger+a^\dagger\DG{5}\right)\ .
\end{equation*}

We can now reduce the numerator, first with the replacement of
$\pslash-\kslash = y\ppslash +x\pslash$.
\begin{equation*}
N^\mu  \Rightarrow \DG{\mu} \left(vV^\dagger + aA^\dagger \right)
\DG{\mu} \left(y \ppslash + x \pslash\right) + \left( Vv^\dagger +
  Aa^\dagger \right) \left(y \ppslash + x \pslash\right) \DG{\mu}\ .
\end{equation*}
We next need the results of these coupling expressions
\begin{equation*}
vV^\dagger + aA^\dagger = Vv^\dagger + Aa^\dagger = \frac{m_\internal -
  m_\external}{M} vv^\dagger - \frac{m_\internal + m_\external}{M}
aa^\dagger = \alpha \ .
\end{equation*}
Thus, we obtain
\begin{align*}
& = 2 p^\mu x \alpha + 2 {p'}^\mu y \alpha\\
& \Rightarrow + \left(p'+p\right)^\mu (x+y) \alpha\\
N^\mu & \Rightarrow - \alpha u i \sigma^{\mu\nu} q_\nu\ .
\end{align*}

Returning to our amplitude, we can now substitute this result
\begin{equation*}
i \amplitude \supset - ieQ_\external \Spinorubar \frac{i
  \sigma^{\mu\nu} q_\nu}{2 m_\external}\left\{ 2 m_\external
  \frac{-i^6 Q_V}{ieQ_\external} g^2 \left({g'}^2 v^\mathrm{b} R\right)
  \int\limits_0^1\dd{u} u \int\frac{\dd[4]{\ell}}{(2\pi)^4}
  \frac{2}{\left(\ell^2 -\Delta +xyq^2\right)^3} \alpha u\right\}
\Spinoru \ ,
\end{equation*}
where the piece in braces is $F_2(q^2)$.  Extracting $F_2(0)$, we find
\begin{equation}
F_2(0) = - \frac{m_\external}{8\pi^2} \frac{Q_V}{Q_\external}
\frac{{g'}^2 v^\mathrm{b} R}{eM} \int\limits_0^1 \dd{u}
\frac{u^2\left((m_\internal-m_\external) vv^\dagger
    -(m_\internal+m_\external) aa^\dagger\right) }{uM^2
  +(1-u)m_\internal^2 +u(u-1)m_\external^2}\ .
\end{equation}

\subsubsection{The Scalar-Scalar Contribution}

For the case $M\neq0$, we must also add the results of
Section~\ref{sec:scalar:charged}, with $v\to V$ and $a\to A$.  This
contribution corresponds to the diagram in
Figure~\ref{fig:vector:wlike:ss}. We obtain
\begin{multline}
F_2(0) = -\frac{m_\external g^2}{8\pi^2} \frac{Q_V}{Q_\external} \frac{1}{M^2}
\int\limits_0^1 \dd{u} u(1-u) \left\{\frac{m_\external\left\{
    (m_\internal-m_\external)^2 vv^\dagger
    +(m_\internal+m_\external)^2 aa^\dagger \right\}
  u}{uM^2+(1-u)m_\internal^2 +u(u-1)m_\external^2}\ + \right. \\ 
\left. \frac{m_\internal \left\{(m_\internal-m_\external)^2 vv^\dagger
      - (m_\internal+m_\external)^2
      aa^\dagger\right\}}{uM^2+(1-u)m_\internal^2
    +u(u-1)m_\external^2} \right\}\ .
\end{multline}
This term is suppressed relative to the vector-vector and
vector-scalar diagrams when the internal and external fermions are
light compared to the vector.

\section{Some Applications: Standard Model Electroweak Contributions}
\label{sec:SM}

We will now apply the results we obtained in the previous section to
the Standard Model electroweak contributions to the \afermion\
of any light fermion (that is, $m_\internal, m_\external \ll M$).  We
will not display the Standard Model Higgs result here, as the
contribution is negligibly small compared to the other contributions.
In this limit, the diagrams containing only unphysical scalars will be
negligible compared to those containing vectors, and will not be
displayed.  The results we obtain here are valid for all of the known
charged fermions except the top quark, whose mass is certainly not
small compared to the vectors.  In that case, we can't ignore the mass
of the top quark compared to the vector (and unphysical scalar)
masses.  We do not deal with that case here.

\subsection{The \Znaught\ Contribution}

In this small fermion mass limit, the \Znaught\ contribution is
\begin{equation*}
\aext^\Znaught(0) = \frac{-m_\external}{8\pi^2} \left(
  \frac{g}{\cos\thetaW} \right)^2 \int\limits_0^1 \dd{u}
\frac{u(u-1)}{(1-u) M_\Znaught^2)} \left\{ 2 m_\external (u-2)
    \left(vv^\dagger + aa^\dagger\right) + 4 m_\external
    \left(vv^\dagger - aa^\dagger\right) \right\}\ .
\end{equation*}
Since $v = (C_R + C_L)/2$ and $a = (C_R - C_L)/2$, we can replace the
vector and axial couplings with the left and right couplings, to place
this result in terms more suitable for Standard Model phenomenology;
\begin{gather}
vv^\dagger + aa^\dagger = \frac{1}{2} \left(C_L C_L^\dagger  + C_R
  C_R^\dagger\right) \to \frac{1}{2} \left(C_L^2 + C_R^2\right)\\
vv^\dagger - aa^\dagger = \frac{1}{2} \left(C_L^\dagger C_R + C_L
  C_R^\dagger\right) \to C_L C_R\ ,
\end{gather}
where the expressions after the arrow are valid when the couplings are
real (that is $CP$ conserving).  Performing the integrations and
substituting for the couplings
\begin{align}
\aext^\Znaught(0) & = \frac{-m_\external}{8\pi^2}
\frac{g^2}{M_\Znaught^2 \cos^2\thetaW} \frac{2}{3} \left\{ m_\external
\left(C_L^2 + C_R^2\right) - 3 m_\internal C_L C_R\right\}\nonumber \\
& \to -\frac{m_\external^2}{8\pi^2} \frac{\Gfermi}{\sqrt{2}}
\frac{16}{3} \left( C_L^2 +C_R^2 - 3 C_L C_R\right)\ ,
\end{align}
where the last line is the limit $m_\internal = m_\external$ (that is,
there are no flavor changing neutral currents), and we have further
used the relations $M_\Znaught \cos\thetaW = M_\Wpart$,
$g^2/8M_\Wpart^2 \equiv \Gfermi/\sqrt{2}$.

\subsection{The \Wpart\ Contribution}

In the small fermion mass limit, we need to calculate only the
vector-vector and vector-scalar diagrams, as mentioned before.  The
vector-vector diagram contributes
\begin{equation*}
F_2(0)^{\Wpart(vv)}(0) = \frac{Q_\Wpart}{Q_\external}
\frac{m_\external}{8\pi^2} \frac{g^2}{4 M_\Wpart^2} \frac{1}{3} \left(
7 m_\external \left(C_L^2 + C_R^2\right) - 18 m_\internal C_L
C_R\right)\ ,
\end{equation*}
where $Q_\external$ is the electric charge of the incoming fermion,
$Q_\Wpart$ is the electric charge of the \Wpart\ entering the
$\Wpart\Wpart\photon$ vertex, and $g' f^{\mathrm{mnr}} = -ieQ_\Wpart$.
The vector-scalar diagram contributes 
\begin{equation*}
F_2^{\Wpart(sv)}(0) = \frac{m_\external}{8\pi^2}
\frac{Q_\Wpart}{Q_\external} \frac{g^2}{4 M_\Wpart^2} \left(
  m_\external \left(C_L^2 + C_R^2\right) - 2 m_\internal C_L C_R
\right)\ ,
\end{equation*}
where ${g'}^2 v^\mathrm{b} R = e M$, and the charges are defined as
above.  Summing these two contributions, we obtain
\begin{equation}
\aext^\Wpart = \frac{m_\external}{8\pi^2}
\frac{Q_\Wpart}{Q_\external} \frac{g^2}{4 M_\Wpart^2} \frac{1}{3} \left(
  10 m_\external \left(C_L^2 + C_R^2\right) - 24 m_\internal C_L C_R
\right)\ ,
\end{equation}

\subsection{The \photon\ Contribution}

To calculate the photon contribution, we use the result calculated for
\Znaught-like gauge bosons, but we drop the contributions from
unphysical scalar modes (as QED is an unbroken gauge theory, there are
no goldstone modes to contaminate our results) and set the gauge mass
to zero.  Furthermore, since QED does not induce flavor changing
interactions, we necessarily have $m_\internal = m_\external$.
Finally, QED has only vector couplings.  Applying these constraints to
the results of Section~\ref{sec:vector:zlike}, we obtain
\begin{equation}
\aext^\photon = \frac{e^2 Q_\external^2}{8 \pi^2} =
\frac{\alphaem}{2\pi} Q_\external^2\ .
\end{equation}

\subsection{Applications to \amuon}

The prototypical example of the use of these formulae is the
calculation of the one loop electroweak corrections to \amuon.  The
following table gives the charges and masses of the Standard Model
weak contributions
\begin{center}
\begin{tabular}{|c|c|c|}
\hline
& \Znaught & \Wpart \\
\hline
$m_\internal$ & $m_\muon$ & $m_\neutrino = 0$\\
$m_\external$ & $m_\muon$ & $m_\muon$\\
$C_L$ & $-\frac{1}{2} + \sin^2\thetaW$ & $\frac{1}{\sqrt{2}}$\\
$C_R$ & $\sin^2\thetaW$ & $0$ \\
\hline
\end{tabular}
\end{center}
Substituting these values into the expressions in the previous
subsections, we find
\begin{gather}
\amuon^\Znaught = -\frac{m_\muon^2 \Gfermi}{8\pi^2\sqrt{2}}
\frac{16}{3} \left(1+2\sin^2\thetaW - 4 \sin^4\thetaW \right)\\
\amuon^\Wpart = \frac{m_\muon^2 \Gfermi}{8\pi^2\sqrt{2}} \frac{10}{3}\\
\intertext{and}
\amuon^\photon = \frac{\alphaem}{2\pi}\ .
\end{gather}
These are in agreement with the Standard Model expectation (see for
example \cite{peskin} or \cite{fls}).

\section{Some Applications: Beyond the Standard Model}
\label{sec:nonSM}

We give here two examples of physics beyond the Standard Model, and
how that physics impacts \amuon.  In particular, we look at a simple
extension of the electroweak symmetry, and the addition of a heavy
Dirac neutrino.

\subsection{Extended Electroweak Symmetry}

Imagine that the gauge group of the electroweak interaction, instead
of $\SUtwo_L \times \Uone_Y$ is actually $\SUtwo_L \times \Uone_1
\times \Uone_2$, where the muon couples to $\Uone_1$ with its
standard hypercharge coupling.  The new massive gauge boson, the
\Zprime, couples to
\begin{equation}
i g' \DG{\mu} \left( \frac{\cphi}{\sphi} Y_2 - \frac{\sphi}{\cphi}
  Y_1 \right)\ ,
\end{equation}
where $\cphi = \cos\phi$, and $\sphi = \sin\phi$ (the mixing angle
between the $\Uone_1$ and $\Uone_2$ gauge fields following spontaneous
symmetry breaking) and gives rise to an \amuon\ contribution that can
be scaled directly from the Standard Model result
\begin{equation}
\amuon^\Zprime = \frac{m_\muon^2 \Gfermi}{8\pi^2 \sqrt{2}}
\frac{4}{3} \frac{M_\Znaught^2}{M_\Zprime^2} \frac{\sphi^2}{\cphi^2}
\sin^2\thetaW\ .
\end{equation}

\subsection{A Heavy Dirac Neutrino}

Imagine adding a heavy neutrino (by which we mean $m_\particle{H} >
M_\Znaught$/2) to the particle content of the Standard Model
(presumably within the content of a fourth generation), and giving it
a small amount of mixing with the muon neutrino,
$\Lambda_{\muon\particle{H}}$.  The vector-vector contribution to
\amuon\ would be given by
\begin{equation*}
F_2(0)^{\particle{H}(vv)} = \frac{m_\muon^2}{8\pi^2} \frac{g^2}{4}
\left|\Lambda_{\muon\particle{H}}\right|^2 \int\limits_0^1 \dd{u}
\frac{u^2(2u+1)}{(1-u) m_\particle{H}^2 + u M_\Wpart^2}\ ,
\end{equation*}
while the vector-scalar contribution is given by
\begin{equation*}
F_2(0)^{\particle{H}(sv)} = \frac{m_\muon^2}{8\pi^2} \frac{g^2}{4} 
\left|\Lambda_{\muon\particle{H}}\right|^2 \int\limits_0^1 \dd{u}
\frac{u^2}{(1-u) m_\particle{H}^2 + u M_\Wpart^2}\ ,
\end{equation*}
and the scalar-scalar contribution is given by
\begin{equation*}
F_2(0)^{\particle{H}(ss)} = \frac{m_\muon^2}{8\pi^2} \frac{g^2}{4}
\left|\Lambda_{\muon\particle{H}}\right|^2 \left(
  \frac{m_\particle{H}}{M_\Wpart} \right)^2 \int\limits_0^1 \dd{u}
\frac{u(u-1)^2}{(1-u) m_\particle{H}^2 + u M_\Wpart^2}\ .
\end{equation*}
Integrating over $u$ and summing the terms, we find the total
contribution to \amuon\ is given by
\begin{equation}
\amuon^\particle{H} = -\frac{m_\muon^2 \Gfermi}{8\pi^2\sqrt{2}}
\left|\Lambda_{\muon\particle{H}}\right|^2
\mathcal{F}(M_\Wpart/m_\particle{H})\ ,
\end{equation}
where 
\begin{equation}
\mathcal{F}(\alpha) = \frac{(1-\alpha)(1+\alpha) (1+ 35\alpha^2
  -40\alpha^4 +10\alpha^6) - 6 \alpha^2(4-3\alpha^2) \ln
  \alpha^2}{1(1-\alpha^2)^4}\ .
\end{equation}
The singularity in this expression at $\alpha = 1$ is not physically
relevant; it is an artifact of taking the zero-width approximation in
our gauge boson propagators.

\section{Conclusions}
\label{sec:conclusion}

In this note, we have presented phenomenologically motivated
calculations of contributions to the $g-2$ of charged fermions,
including both neutral and charged scalars and vectors.  Our
expressions reproduce the Standard Model electroweak contributions to
\amuon\ in the appropriate mass limits, and are flexible enough to
allow us to handle many scenarios of new physics beyond the Standard
Model.  The most obvious extensions of the results presented here are
the following:
\begin{enumerate}
\item We derived our results involving gauge bosons in Feynman Gauge;
  a useful extension of the work would involve performing the
  calculations in a generalized $R_\xi$ gauge.  
\item The results were derived in all cases under the assumption of
  narrow boson widths, that is, $\Gamma_X / M_X \ll 1$.  This is not
  necessarily a defensible assumption in some models, or even in
  special corners of parameter space in models where this is generally
  a valid choice.  Care, then, must be taken when applying the results
  given here to models with wide bosons.
\end{enumerate}
Of course, more involved extensions are also possible.  In particular,
we have calculated the one-loop amplitude for $\particle{f} \to
\particle{f}\photon$, but there is no real impediment to extending our
results to processes such as $\particle{f} \to \particle{f'}\photon$.
The calculation of these amplitudes is necessary for the study of, for
example, lepton number violating processes such as $\muon \to
\electron\photon$, or quark-quark transitions such as $\qbottom \to
\qstrange\photon$.  Calculations of such processes exist in many
places in the literature; in particular, we cite \cite{Bjorken:1977br}
for a general analysis of the $\muon \to \electron \photon$
transition. 

\vspace{24pt}
\begin{center}\textbf{Acknowledgments}\end{center} 
\vspace{12pt} 

Thanks go to C.~H\"{o}lbling, M.~Popovic, T.~Rador, and E.~H.~Simmons,
for useful discusions.  E.~H.~Simmons deserves special thanks for
insightful comments on the manuscript that greatly improved the
readability and utility of the document.  Thanks also go to N.~Rius
for bringing a sign error in the derivations of the scalar
contributions to our attention.  This work was supported in part by
the Department of Energy under grant DE-FG02-91ER40676, the National
Science Foundation under grant PHY-0074274, and by the Radcliffe
Institute for Advanced Study.

\bibliographystyle{apsrev}
\bibliography{note}

\begin{thebibliography}{10}
\expandafter\ifx\csname bibnamefont\endcsname\relax
  \def\bibnamefont#1{#1}\fi
\expandafter\ifx\csname bibfnamefont\endcsname\relax
  \def\bibfnamefont#1{#1}\fi
\expandafter\ifx\csname url\endcsname\relax
  \def\url#1{\texttt{#1}}\fi
\expandafter\ifx\csname urlprefix\endcsname\relax\def\urlprefix{URL }\fi
\providecommand{\bibinfo}[2]{#2}
\providecommand{\eprint}[2][]{\url{#2}}

\bibitem{gauge}
\bibinfo{author}{\bibfnamefont{K.~R.} \bibnamefont{Lynch}}
  (\bibinfo{year}{2001}), \eprint{BUHEP-01-16}.

\bibitem{Brown:2001mg}
\bibinfo{author}{\bibfnamefont{H.~N.} \bibnamefont{Brown}} \emph{et~al.}
  (\bibinfo{collaboration}{Muon g-2}), \bibinfo{journal}{Phys. Rev. Lett.}
  \textbf{\bibinfo{volume}{86}}, \bibinfo{pages}{2227} (\bibinfo{year}{2001}),
  \eprint{hep-ex/0102017}.

\bibitem{Jackiw:1972jz}
\bibinfo{author}{\bibfnamefont{R.}~\bibnamefont{Jackiw}} \bibnamefont{and}
  \bibinfo{author}{\bibfnamefont{S.}~\bibnamefont{Weinberg}},
  \bibinfo{journal}{Phys. Rev.} \textbf{\bibinfo{volume}{D5}},
  \bibinfo{pages}{2396} (\bibinfo{year}{1972}).

\bibitem{fls}
\bibinfo{author}{\bibfnamefont{K.}~\bibnamefont{Fujikawa}},
  \bibinfo{author}{\bibfnamefont{B.~W.} \bibnamefont{Lee}}, \bibnamefont{and}
  \bibinfo{author}{\bibfnamefont{A.~I.} \bibnamefont{Sanda}},
  \bibinfo{journal}{Phys. Rev. D} \textbf{\bibinfo{volume}{6}},
  \bibinfo{pages}{2923} (\bibinfo{year}{1972}).

\bibitem{Leveille:1978rc}
\bibinfo{author}{\bibfnamefont{J.~P.} \bibnamefont{Leveille}},
  \bibinfo{journal}{Nucl. Phys.} \textbf{\bibinfo{volume}{B137}},
  \bibinfo{pages}{63} (\bibinfo{year}{1978}).

\bibitem{peskin}
\bibinfo{author}{\bibfnamefont{M.~E.} \bibnamefont{Peskin}} \bibnamefont{and}
  \bibinfo{author}{\bibfnamefont{D.~V.} \bibnamefont{Schroeder}},
  \emph{\bibinfo{title}{An Introduction to Quantum Field Theory}}
  (\bibinfo{publisher}{Addison-Wesley Publishing Company},
  \bibinfo{address}{Reading, MA}, \bibinfo{year}{1995}).

\bibitem{cheng}
\bibinfo{author}{\bibfnamefont{T.-P.} \bibnamefont{Cheng}} \bibnamefont{and}
  \bibinfo{author}{\bibfnamefont{L.-F.} \bibnamefont{Li}},
  \emph{\bibinfo{title}{Gauge Theory of Elementary Particle Physics}}
  (\bibinfo{publisher}{Oxford University Press}, \bibinfo{address}{New York,
  NY}, \bibinfo{year}{1994}).

\bibitem{zuber}
\bibinfo{author}{\bibfnamefont{C.}~\bibnamefont{Itzykson}} \bibnamefont{and}
  \bibinfo{author}{\bibfnamefont{J.-B.} \bibnamefont{Zuber}},
  \emph{\bibinfo{title}{Quantum Field Theory}} (\bibinfo{publisher}{McGraw-Hill
  Book Company}, \bibinfo{address}{New York, NY}, \bibinfo{year}{1980}).

\bibitem{quigg}
\bibinfo{author}{\bibfnamefont{C.}~\bibnamefont{Quigg}},
  \emph{\bibinfo{title}{Gauge Theories of the Strong, Weak, and Electromagnetic
  Interactions}} (\bibinfo{publisher}{Addison-Wesley Publishing Company},
  \bibinfo{address}{Reading, MA}, \bibinfo{year}{1983}).

\bibitem{Bjorken:1977br}
\bibinfo{author}{\bibfnamefont{J.~D.} \bibnamefont{Bjorken}},
  \bibinfo{author}{\bibfnamefont{K.}~\bibnamefont{Lane}}, \bibnamefont{and}
  \bibinfo{author}{\bibfnamefont{S.}~\bibnamefont{Weinberg}},
  \bibinfo{journal}{Phys. Rev.} \textbf{\bibinfo{volume}{D16}},
  \bibinfo{pages}{1474} (\bibinfo{year}{1977}).

\end{thebibliography}

\end{document}